\documentclass[11pt]{article}
\pdfoutput=1  
\usepackage{graphicx}

\usepackage{appendix}

\usepackage{amssymb}
\usepackage{amsmath}
\usepackage{amsfonts}
\usepackage{upgreek}
\usepackage{latexsym}
\usepackage{slashed,upgreek,amscd,cancel,tensor}
\usepackage{adjustbox}
\usepackage{epsfig}
\usepackage{cite}
\usepackage{hyperref}
\usepackage{amsfonts}
\usepackage{booktabs}
\usepackage{siunitx}
\usepackage{caption}
\usepackage{float}
\usepackage{color}

\numberwithin{equation}{section}

\definecolor{MyBlue}{rgb}{0.15,0.15,0.70}

\hypersetup{
colorlinks=true,
citecolor=MyBlue,
linkcolor=MyBlue,
urlcolor=MyBlue
}

\input epsf
\newcommand\Ups{\Upsilon}

\newcommand{\be}{\begin{equation}}
\newcommand{\ee}{\end{equation}}
\newcommand{\beq}{\begin{equation}}
\newcommand{\eeq}{\end{equation}}
\newcommand{\bea}{\begin{eqnarray}}
\newcommand{\eea}{\end{eqnarray}}

\newcommand{\DDt}{\alpha}

\usepackage[stable]{footmisc}
\def\d{\delta}

\def\dkmu2{\delta K_{\mu \nu}\delta K^{\mu \nu}}
\def\pmu2{  \phi_{\mu \nu}\phi^{\mu \nu}}

\newcommand{\fg}{f_{\rm eff}}

\renewcommand\[{\left[}

\newcommand\ees{\end{eqnarray}}
\newcommand\bees{\begin{eqnarray}}

\newcommand\alphaDI{\alpha_{\text{D},I}}
\newcommand\alphaCI{\alpha_{\text{C},I}}

\newcommand\alphaDc{\alpha_{\text{D},c}}
\newcommand\alphaCc{\alpha_{\text{C},c}}
\newcommand\alphaD{\alpha_{\text{D}}}
\newcommand\alphaC{\alpha_{\text{C}}}
\newcommand\alphaB{\alpha_{\text{B}}}
\newcommand\alphaM{\alpha_{\text{M}}}
\newcommand\alphaK{\alpha_{\text{K}}}
\newcommand\alphaT{\alpha_{\text{T}}}

\newcommand\tbeta{\beta}
\newcommand\gammac{\gamma_c}

\newcommand\timedep{\Gamma}
\newcommand\betaB{\beta_{\text{B}}}

\newcommand\wDE{w}

\newcommand\GG{G} 

\begin{document}

\hfill CERN-PH-TH-2015-223
\vspace{0.5cm}

\begin{center}
\LARGE{\textbf{Effective Theory of Dark Energy \\at Redshift Survey Scales }} \\[1cm] 
 
\Large{J\'er\^ome Gleyzes$^{\rm a,b}$,  David Langlois$^{\rm c}$, \\[0.1cm]
Michele Mancarella$^{\rm a,b,d}$ and Filippo Vernizzi$^{\rm a,d}$}
\\[0.7cm]

\small{
\textit{$^{\rm a}$ Institut de physique th\' eorique, Universit\'e  Paris Saclay \\ [0.05cm]
CEA, CNRS, 91191 Gif-sur-Yvette, France}}

\vspace{.2cm}

\small{
\textit{$^{\rm b}$ Universit\'e Paris Sud, 15 rue George Cl\'emenceau, 91405,  Orsay, France}}

\vspace{.2cm}

\small{
\textit{$^{\rm c}$  APC, (CNRS-Universit\'e Paris 7), 10 rue Alice Domon et L\'eonie Duquet, 75205 Paris, France \\ 
}}
\vspace{.2cm}

\small{
\textit{$^{\rm d}$  Physics Department, Theory Unit, CERN, CH-1211 Gen\`eve 23, Switzerland \\ 
}}
\vspace{.2cm}

\vspace{0.5cm}
\today

\end{center}

\vspace{2cm}

\begin{abstract}
We explore the phenomenological consequences of general late-time modifications of gravity in the quasi-static approximation, in the case where cold dark matter is non-minimally coupled to the gravitational sector.
Assuming spectroscopic and photometric surveys
with configuration parameters similar to those of the Euclid mission, we  derive 
constraints on our effective description from three observables:   the galaxy power spectrum in redshift space, 
tomographic weak-lensing shear power spectrum and the correlation spectrum between the integrated Sachs-Wolfe effect and the galaxy distribution. 
In particular, with $\Lambda$CDM as fiducial model and a specific choice for the time dependence of our effective functions, we perform a Fisher matrix analysis and find that the unmarginalized $68\%$ CL errors on 
the parameters describing the modifications of gravity are of  order  $\sigma\sim10^{-2}$--$10^{-3}$. We also consider two other fiducial models.
A nonminimal coupling of CDM enhances the effects of modified gravity
and reduces the above statistical errors accordingly.
In all cases, we find that the parameters are highly degenerate, which prevents the inversion of the Fisher matrices. Some of these degeneracies can be broken by combining all three observational probes. 

\end{abstract}

\newpage 
\tableofcontents

\vspace{.5cm}
\newpage

\section{Introduction}

The recent measurements of the cosmic microwave background (CMB) anisotropies, performed  by the WMAP and Planck satellites,  have  significantly improved our knowledge on the content of the universe and  
on the initial conditions of cosmological perturbations. A similar progress is expected from the next generation of galaxy surveys concerning the properties of dark energy or, possibly,  modifications of general relativity on cosmological scales. Indeed, even if the  CMB is useful to constrain dark energy through the integrated Sachs-Wolfe (ISW) effect and gravitational lensing, these effects are ultimately related to  the impact of dark energy on the late-time evolution of structures. Probing directly these large scale structures is thus thought to be  the most promising source of  information on the origin of the current acceleration. 

Since no compelling model of dark energy has emerged from theoretical investigations, it is appropriate to resort to a description that encodes a wide range of physical effects with a limited number of theoretically motivated parameters, in order to compare deviations from the standard $\Lambda$CDM scenario with cosmological observations on linear scales. For single-field dark energy models in the presence of universally coupled matter fields, this research program has been  initiated by the effective theory of dark energy  recently proposed in Refs.~\cite{Gubitosi:2012hu,Bloomfield:2012ff,Gleyzes:2013ooa}, inspired by the so-called effective field theory of inflation \cite{Creminelli:2006xe,Cheung:2007st} and of minimally coupled dark energy \cite{Creminelli:2008wc}. Another model-independent framework that has been developed with the same motivations is the Parameterized Post-Friedmann approach \cite{Baker:2012zs,Ferreira:2014mja}. In the effective theory of dark  energy, the quadratic action describing linear perturbations of  single-field models  belonging to Horndeski theories is characterized by four free functions of time \cite{Gleyzes:2013ooa,Bloomfield:2013efa,Bellini:2014fua,Gleyzes:2014rba}, while a fifth function must be introduced to describe theories beyond Horndeski \cite{Gleyzes:2014dya,Gleyzes:2014qga}. 
The power and efficiency of this formalism  has just started to be exploited. For instance, it has been applied  to explore and forecast the phenomenology of dark energy and modified gravity in  \cite{Piazza:2013pua,Ade:2015rim,Tsujikawa:2015mga,Brax:2015pka}
(see also \cite{Bellini:2015wfa,Bellini:2015oua} for some nonlinear aspects).

Recently, in Ref.~\cite{Gleyzes:2015pma}, we extended this unifying treatment  to allow for distinct conformal and disformal couplings of matter species to the gravitational sector.\footnote{A treatment of single-field dark energy coupled to CDM in the context of the Parameterized Post-Friedmann framework can be found in \cite{Skordis:2015yra}.} We focused on Horndeski-like models, i.e.~those whose quadratic action has the same structure as Horndeski theories,\footnote{Note that although Horndeski theories are generically unstable under quantum corrections \cite{Nicolis:2008in}, an example of a radiatively stable subclass of Horndeski theories where all the operators of action (2.3) can be relevant has been proposed in \cite{Pirtskhalava:2015nla}, based on weakly broken galileon invariance, and applied to inflation in \cite{Pirtskhalava:2015zwa}.} although the full action can be different.
This is a rather natural extension given that a modification of the gravitational sector can often be interpreted as a direct coupling of matter to a fifth force exchanged by the scalar, in the frame where the scalar and the gravitational fluctuations are demixed---the so-called Einstein frame. 
Together with the four functions describing the gravitational quadratic action, each matter species is now characterized by two new functions parametrizing their conformal and disformal couplings to the gravitational metric.  However, as reviewed in Sec.~\ref{sec2},  the structure of the full action remains invariant under conformal and disformal  transformations of the gravitational metric itself.
Taking into account this freedom, which allows for instance to choose a frame where one of the species is minimally coupled, one eventually 
finds that the whole system depends  on a total of $2 (N_S + 1)$ independent functions of time, where  $N_S $ is the number of  matter species. In this context, 
the conditions for stability (i.e.~the absence of ghostlike and gradient instabilities) can be generalized to  any frame (see Sec.~\ref{sec2}).

In this article we go one step further and explore the constraining power of future large scale structure surveys on the deviations from the standard $\Lambda$CDM scenario, expressed in terms of the parameters of the effective theory of dark energy proposed in \cite{Gleyzes:2015pma}.
Specifically, we will consider a simple scenario where the gravitational sector is described by Horndeski-like models while, in the matter sector, cold dark matter (CDM) is nonminimally coupled to gravity. This extends  to a much broader spectrum of gravitational theories previous studies of coupled dark energy, with conformal \cite{Damour:1990tw,Amendola:1999er} (see also \cite{Amendola:2012ys} and references therein) and disformal (see  e.g.~\cite{Koivisto:2008ak,Zumalacarregui:2010wj,Koivisto:2012za,vandeBruck:2012vq,Zumalacarregui:2012us,Brax:2013nsa,Brax:2014vva,Sakstein:2014isa,vandeBruck:2015ida,Koivisto:2015mwa}) couplings. 

The equations of motion for the linear perturbations  in the presence of modified gravity and nonmininally coupled CDM,
  derived in \cite{Gleyzes:2015pma}, are reviewed in Sec.~\ref{sec3}, where we assume the quasi-static approximation. As shown in \cite{Sawicki:2015zya}, this approximation should be reliable  for surveys such as Euclid  as long as the sound speed exceeds $10 \%$ of the speed of light, i.e.~$c_s \gtrsim 0.1$. 
In particular, we will consider  the extreme quasi-static limit, i.e.~the limit $k\to \infty$, of the dynamics.
In such a regime the linear growth of  matter (both for baryons and CDM)   remains scale-independent as in $\Lambda$CDM. Modifications of gravity and the nonminimal coupling to CDM are encoded in the time dependence of the gravitational couplings in the ``Poisson'' equations for the metric potentials, which are different for baryons and CDM. As explained in Sec.~\ref{sec3}, this time dependence modifies the growth rate of structures and the lensing potential, which in turn affect, respectively, the redshift-space distortions and the weak-lensing cosmic shear.

In Sec.~\ref{sec:timepar} we introduce the details of  our 
parametrization, in particular concerning the time dependence of the parameters characterizing the modifications of gravity. We consider three fiducial models: a minimal $\Lambda$CDM model, a braiding model and a model with an active nonminimal coupling of CDM.
In Sec.~\ref{sec5} we perform a Fisher matrix analysis
based on future photometric and spectroscopic data with configuration parameters close to those of the Euclid mission \cite{Laureijs:2011gra,Amendola:2012ys} as an example.
We focus on the two-point statistics and consider the galaxy power spectrum in redshift space for the spectroscopic data, 
the projected weak-lensing shear power spectrum for the photometric data as well as the correlation between the ISW effect in the CMB temperature and the photometric galaxy distribution. 
The derived constraints are discussed in Sec.~\ref{sec6}, together with the involved degeneracies. It should be mentioned that other approaches have been developed to study in a general and model-independent way the impact of modified gravity on cosmological observables, together with the involved degeneracies,  e.g.~on the growth rate of fluctuations \cite{Baker:2013hia}  (see also \cite{Burrage:2015lla,Taddei:2014wqa}) or on the weak lensing  \cite{Leonard:2015hha}.

In Sec.~\ref{conclusion} we summarize our results and draw conclusions.
Details on the parametrization and the choice of background cosmological parameters are given in the App.~A, while in App.~B, we discuss the frame dependence of the evolution equations of matter.

\section{Model and main equations}
\label{sec2}
In this section, we introduce our general formalism and then focus on the specific model at the core of the present work. The first subsection, which is mainly a review of our recent paper \cite{Gleyzes:2015pma} and previous works, can be skipped by the reader mostly interested in our phenomenological model and forecasts for the parameter constraints. 
The model that we are specifically studying in the rest of this paper is described in the second subsection.

\subsection{Effective description of the gravitational and matter sectors}
We start by summarizing  the effective approach of dark energy introduced and developed in Refs.~\cite{Gubitosi:2012hu,Gleyzes:2013ooa,Gleyzes:2015pma} (see e.g.~\cite{Piazza:2013coa,Gleyzes:2014rba} for reviews). 
The gravitational sector is assumed to consist of a four-dimensional metric $g_{\mu \nu}$ and of  a scalar field $\phi$. In order to treat simultaneously a wide range of models, it is very convenient to ``hide'' the scalar field in the metric, by choosing the constant-time hypersurfaces to coincide with the  uniform scalar field hypersurfaces. 
In this gauge, referred to  as unitary gauge, the metric  can be written in the ADM form \cite{Arnowitt:1962hi}, 
  \be
\label{ADM}
ds^2=-N^2 dt^2 +{h}_{ij}\left(dx^i + N^i dt\right)\left(dx^j + N^j dt\right) \, ,
\ee
where $N$ is the lapse function,   $N^i$ the shift vector and $h_{ij}$ the three-dimensional  spatial metric.

In  unitary gauge, a generic gravitational action  can be written in terms of  geometric quantities that are invariant under spatial diffeomorphisms, namely in terms of 
the lapse $N$, the 3d Ricci tensor $R_{ij}$ of the constant time hypersurfaces, as well as  their extrinsic curvature $K_{ij}$,
with components
 \be
\label{extrinsic_ADM}
K_{ij} = \frac{1}{2N} \big(\dot h_{ij} - D_i N_j - D_j N_i \big) \;,
\ee
where a dot stands for a time derivative with respect to $t$, and $D_i$ denotes the covariant derivative associated with the spatial metric $h_{ij}$ (spatial indices are lowered or raised via  the  metric $h_{ij}$).

 The generalized Friedmann equations are then obtained by varying the specialization of the action to a homogeneous FLRW 
 (Friedmann-Lema\^itre-Robertson-Walker)  spacetime, endowed with the metric $ds^2=-dt^2+a^2(t)d\vec {x}^2$\,. 
The dynamics of the linear perturbations is governed by the quadratic action, obtained by a perturbative expansion of the original action. 

In this paper, we will consider a very large class of models, which includes all Horndeski theories, for which the quadratic action can be written in the form~\cite{Gleyzes:2013ooa,Bloomfield:2013efa,Bellini:2014fua,Gleyzes:2014rba}\footnote{Together with the operator $\alpha_{\rm H} \delta N R$, this is the most general quadratic action for linear perturbations about a homogeneous and isotropic spacetime that does not induce higher derivatives in the equation of motion of the linearly propagating scalar degree of freedom. In consistent effective theories, higher time derivatives are not forbidden but are suppressed by positive powers of the ratio between the energy and the cutoff scale (see e.g.~\cite{Weinberg:2008hq,Burgess:2014lwa}). Thus, at energies much smaller than the cutoff their effect can be neglected without loss of generality. Higher spatial derivatives are not necessarily suppressed and may dominate the dispersion relation, such as in the Ghost Condensate theory \cite{ArkaniHamed:2003uy}. 
In this case, higher spatial gradients  become relevant, and can easily be included in our formalism, but begin operating at very short distances \cite{ArkaniHamed:2005gu,Creminelli:2008wc}, typically shorter than the cosmological ones.}
\be
\begin{split} 
\label{S2}
 S_{\rm g}^{(2)}=  \int d^3x dt \,a^3  \frac{M^2}{2}   \bigg[ & \delta K^i_j \delta K^j_i-\delta K^2 + R \delta N+(1+\alphaT) \, \delta_2 \Big(   {\sqrt{h}}R/{a^3 }\Big)    \\
 &   +  \alphaK H^2 \delta N^2 + 4 \alphaB H \delta K \delta N  \bigg]   \, ,
 \end{split}
 \ee
 where $M$, $\alphaT$, $\alphaB$ and $\alphaK$ are four time-dependent functions 
and $\delta_2$ denotes the second order term in a perturbative expansion. $H\equiv \dot a/a$ is the Hubble parameter. 
We have not included  irrelevant terms that vanish when adding the matter action and imposing the background equations of motion. Note that (\ref{S2}) does not include the models beyond Horndeski \cite{Gleyzes:2014dya} for which the coefficient of the term $R\,\delta N$ differs from $1$, the difference defining a new parameter $\alpha_{\rm H}$\cite{Gleyzes:2014rba}.

General relativity corresponds to the particular case where $\alphaT=\alphaB=\alphaK=0$ and $M=M_{\rm Pl}$. In general, the above quadratic action contains not only two tensor modes, as in general relativity, but a scalar mode as well. The coefficient in front of the tensor kinetic term is $M^2$ and, by analogy with general relativity, $M$ can be identified with an effective Planck mass. If $M$ depends on time, it is convenient to introduce the related parameter
\beq
\label{alphaM}
\alphaM\equiv \frac{1}{ H}\frac{d \ln M^2}{ dt}\,.
\eeq
The parameter $\alphaT$ appears in  the gradient term of the tensor modes and is thus directly related to the tensor propagation speed, namely 
\beq
\label{cTsq}
c_T^2\equiv 1+\alphaT\,.
\eeq
The stability of the tensor modes is ensured by requiring $M^2>0$ (absence of ghosts) and 
$\alphaT > -1$ (absence of gradient instabilities).\footnote{As shown in \cite{Gleyzes:2015pma,Creminelli:2014wna} and reviewed below,  the propagation speed for gravitons can be set  to unity by a convenient disformal transformation (only ratios between sound speeds are invariant and thus meaningful physical quantities). It is thus not a priori pathological to have $c_T^2>1$ in a generic frame
and we will not impose any upper bound on $c_T$ as a condition for the viability of the theory.  
A propagation speed for gravitons {\em smaller} than that of the other particles is instead very tightly constrained at high energy by  cosmic rays observations \cite{Moore:2001bv}. We have not taken this bound into account in our analysis, since it concerns the speed of gravitational waves at wavelengths much shorter than the cosmological ones.}

Keeping in mind that the lapse perturbation is analogous, in the ADM language, to the time derivative of the scalar perturbation, one observes that  the parameter $\alphaK$ is related to the coefficient of the kinetic scalar term. It is thus present for simple quintessence models. Finally, the coefficient $\alphaB$ characterizes the mixing between the scalar  and tensor kinetic terms, sometimes called ``braiding''. 
In contrast with the tensor modes, the full dynamics of the  scalar mode depends on the matter action as well, and the discussion on the scalar stability conditions  thus needs to be postponed 
until after the introduction of the matter action  below.

The remarkably simple form of the quadratic action (\ref{S2}) holds only in  the unitary gauge. However, it is  straightforward to derive the quadratic action in an arbitrary gauge, by simply performing a time reparametrization of the form
\be
t\rightarrow \phi= t +\pi (t,{\mathbf x}) \;,
\ee
where the unitary time becomes a four-dimensional scalar field. 
The scalar degree of freedom of the gravitational sector thus reappears explicitly in the form of the scalar perturbation $\pi$.

A matter species can be either minimally or nonminimally coupled to the gravitational metric $g_{\mu\nu}$. In the latter case, it is often assumed that matter is  minimally coupled to some effective metric $\tilde g_{\mu\nu}$, which depends on $g_{\mu\nu}$ and on the scalar field $\phi$. We will adopt this type of nonminimal coupling in the following and consider a matter action of the form
\beq
S_{\rm m}=S_{\rm m}[\psi_{\rm m}, \tilde g_{\mu\nu}]\,,
\eeq
with
\beq
\label{transf_cd}
\tilde g_{\mu\nu}= C(\phi) g_{\mu\nu}+D(\phi)\partial_\mu\phi\partial_\nu\phi\,.
\eeq

The initial gravitational metric $g_{\mu\nu}$ being  somewhat arbitrary in general, one has the freedom to choose  the metric $\tilde g_{\mu\nu}$ 
as the new gravitational metric. Remarkably, the quadratic action (\ref{S2}) remains of the same form \cite{Bettoni:2013diz,Gleyzes:2015pma},
\footnote{In the presence of the operator proportional to $\alpha_{\rm H}$ \cite{Gleyzes:2013ooa,Gleyzes:2014rba} describing linear perturbations in the theories beyond Horndeski proposed in \cite{Gleyzes:2014dya,Gleyzes:2014qga}, the structure of the Lagrangian remains invariant under the transformation \eqref{transf_cd} even   
 if the disformal function $D$ depends on $ (\partial \phi)^2 $  as well
 \cite{Gleyzes:2014dya} (see also \cite{Tsujikawa:2015upa} for a recent study).
} 
with new parameters defined as
\be
 \tilde{M}^2 =\frac{M^2 }{ C\sqrt{1+\alphaD}}\, \label{Mtilde}
\ee
and\footnote{Here we correct a typo in the expression for $\tilde \alpha_{\rm K}$ in eq.~(2.45) of the arXiv version of Ref.~\cite{Gleyzes:2015pma}.}
\be
\begin{split} 
\label{alphatilde}
\tilde{\alpha}_{\rm K} &=\frac{\alphaK+12\alphaB[\alphaD+(1+\alphaD)\alphaC]-6[ \alphaD+(1+\alphaD)\alphaC]^2 +3\Omega_{\rm m}\alphaD}{(1+\alphaC)^2(1+\alphaD)^2} \;, \\ \tilde{\alpha}_{\rm B}&= \frac{1+\alphaB}{(1+\alphaC)(1+\alphaD)}-1 \;,  \\
\tilde{\alpha}_{\rm M} &= \frac{\alphaM- 2 \alphaC }{1+\alphaC} - \frac{\dot \alpha _{\rm D}}{2 H (1+\alphaD) (1+\alphaC)} \;, \\
\tilde{\alpha}_{\rm T}&=(1+\alphaT)(1+\alphaD)-1\;,
\end{split}
\ee
where 
\be
\label{defalphas}
 \alphaC \equiv \frac{\dot C}{2 H C} \, , \qquad \alphaD \equiv \frac{D}{ C-D}\,.
 \ee
Given a single species of matter, one can thus always work in the frame where this species is minimally coupled. If there are several matter species, this is possible only in the case of universal coupling, i.e.~if all species are coupled to gravity via the same effective metric. By contrast, for species with different couplings, one cannot  find a frame where all of them are minimally coupled. It remains however possible to choose a frame where one of the species is minimally coupled, even if the others are not.\footnote{The situation simplifies during inflation, when the couplings to matter can be ignored. In this case,  without loss of generality one can always go to a frame where $\tilde \alpha_{\rm M} = \tilde \alpha_{\rm T}=0$, corresponding to the standard time-independent Planck mass and unity speed of propagation for gravitons. In this frame one then recovers the standard inflationary predictions \cite{Creminelli:2014wna}.}

The sum of the gravitational and matter actions at quadratic order yields the dynamics of the scalar mode, as mentioned earlier. As shown in \cite{Gleyzes:2015pma}, the kinetic term of the scalar mode is proportional to the combination
\be
\DDt \equiv \alphaK + 6 \alphaB^2 +3 \sum_I \alphaDI \, \Omega_I\;, \label{alpha_def}
\ee
where
\be
\label{defs_Omega}
\Omega_{I}\equiv \frac{\rho_{I}}{3H^2M^2}  \;, 
\ee
while its propagation speed is given by
\be
\label{ss}
c_s^2 = -\frac{2}{\DDt}\bigg\{ (1+\alphaB) \bigg[ \frac{\dot H}{H^2} -\alphaM + \alphaT  +  \alphaB (1+\alphaT)\bigg] + \frac{\dot \alpha_{\rm B}}{H} +\frac32 \sum_I  \Big[ 1+ (1+\alphaDI) w_I \Big] \Omega_I\bigg\} \;.
\ee
The stability conditions for the scalar mode,
\beq
\DDt \ge 0\,,\qquad c_s^2 \ge0\,,
\eeq
involve all the modified gravity parameters, as well as the matter disformal couplings.

\subsection{Baryon-CDM model}

In our model, the coupling of CDM to the gravitational sector is different from that of the other species (baryons, photons and neutrinos). In the following, for simplicity, we choose to work in the frame where the other species are minimally coupled and 
assume that the original metric $g_{\mu\nu}$ corresponds to this frame (if not, one just needs to apply the above metric transformation). 
We then assume that the coupling of CDM to gravity and dark energy  is characterized by an effective metric 
of the form 
\be
 \check{g}^{(c)}_{\mu\nu}\equiv C_c(\phi) g_{\mu\nu}+D_c(\phi) \partial_\mu \phi \partial_\nu \phi\,,
\ee
from which one can define, in analogy with (\ref{defalphas}), the conformal and disformal parameters
\be
\label{defalphac}
 \alpha_{\text{C},c} \equiv \frac{\dot C_c}{2 H  C_c} \, , \qquad \alpha_{\text{D},c}\equiv \frac{D_c}{ C_c-D_c}\,.
 \ee
We  ignore the photon and neutrino cosmological fluids, as we are interested in late-time cosmology where their effects are negligible.
 
 The equations of motion for the matter species follow from the conservation, or non-conservation, of their respective energy-momentum tensor. Since baryons are minimally coupled, their energy-momentum tensor is conserved as usual, i.e.
\be
\nabla_\mu T_{(b)}{}_{ \ \nu}^\mu =0  \label{evol_matterb}\,.
\ee
By contrast, the CDM energy-momentum tensor is not conserved, but instead satisfies the equation
\be
\nabla_\mu T_{(c)}{}_{ \ \nu}^\mu + Q_c \partial_\nu \phi=0  \label{evol_matterc} 
\ee
with 
\be
\label{def_Qc}
Q_c \equiv - \frac{ C_c' }{2 C_c} T_{(c)} - \frac{D_c'}{2C_c} T_{(c)}^{\mu \nu} \partial_\mu \phi \partial_\nu \phi + \nabla_\mu \left(T_{(c)}^{\mu \nu} \partial_\nu \phi \frac{D_c}{C_c} \right) \;,
\ee
where a prime denotes a derivative with respect to $\phi$. Like the usual conservation equation, this equation can be derived by simply using the invariance of the matter action under arbitrary diffeomorphisms.

The background evolution equations for the baryon and CDM fluids follow directly from (\ref{evol_matterb}) and (\ref{evol_matterc}). On a FLRW background, the definition of $Q_c$, eq.~\eqref{def_Qc},  reduces to
\be
\bar{Q}_c= \frac{H \rho_c}{1+\alpha_{\text{D},c}}    \left\{     \alpha_{\text{C},c} +   \alpha_{\text{D},c} \left(3 +  \frac{\dot\rho_c}{H \rho_c}  \right)
+ \frac{\dot\alpha_{\text{D},c}}{2H (1+\alpha_{\text{D},c})}\right\}   \,. \label{barQ}
\ee
Substituting the above expression into eq.~\eqref{evol_matterc}, one finds that the homogeneous fluid  equations can be written in the form 
\begin{align}
\dot \rho_b + 3 H \rho_b &= 0 \;, \label{cons_1}\\
\dot \rho_c + 3 H (1- \gamma_c) \rho_c & = 0\;, \label{cons_2} 
\end{align}
where  the coupling parameter $\gammac$ is given by\footnote{Taking into account eq.~\eqref{cons_2} one finds that $ \bar Q_c =3H \rho_c \gamma_c$.}
\be
\label{gammac}
\gammac=\frac13 \alphaCc+\frac{\dot \alpha_{\text{D},c}}{6H(1+\alphaDc)}\,.
\ee

Expressed in terms of the energy density fractions defined in (\ref{defs_Omega}),
the evolution equations for the baryon and CDM energy densities, (\ref{cons_1}) and (\ref{cons_2}), become
\begin{align}
\dot\Omega_b &=-H \bigg( 3 + 2 \frac{\dot H}{H^2} + \alphaM\bigg) \Omega_b \,, \label{Omegab}\\
\dot  \Omega_c&= - H \bigg( 3 + 2 \frac{\dot H}{H^2} -3\gammac+ \alphaM\bigg) \Omega_c  \,. \label{Omegac}
\end{align}
The presence of the coefficient $\alphaM$ is due to the fact that  the mass $M$, which appears in the definition (\ref{defs_Omega}), can be time-dependent. 

The evolution of the Hubble parameter is usually determined by the Friedmann equations. In the present work where dark energy remains unspecified at the background level, one can alternatively assume some specific evolution $H=H(t)$ and infer
from it the dark energy background components. This means that the 
Friedmann equations, written in the form 
\be
H^2 = \frac{1}{3 M^2} \left( \rho_{\rm m} + \rho_{\rm DE}\right) \;, \qquad \dot H = - \frac{1}{2 M^2} \left[\rho_{\rm m} + (1+w_{\rm DE}) \rho_{\rm DE} \right]\,,\qquad  \rho_{\rm m} \equiv \rho_{ b}+\rho_{ c}\;,
\ee
are treated as definitions of the energy density for dark energy, $\rho_{\rm DE}$, and of its equation of state parameter, $w_{\rm DE}$, namely 
\be
\label{eos}
 \rho_{\rm DE}  \equiv  3 M^2 H^2  - \rho_{\rm m} \;, \qquad
 w_{\rm DE}   \equiv  \frac{- \frac{2}{3} \frac{\dot H}{H^2} - 1}{1- \Omega_{\rm m}} \;, 
\ee
where
\be
\label{Om}
\Omega_{\rm m} \equiv \Omega_{ b}+\Omega_{ c} \,.
\ee

Given  some prescription
for the time-dependent functions $H=H(t)$, $\alphaM(t)$ and $\gammac(t)$, the evolution of $\Omega_{b}$ and $\Omega_{c}$ can be determined in terms of their present values   $\Omega_{b,0}$ and $\Omega_{c,0}$. This  will be done explicitly in Sec.~\ref{timedep}.

\section{Linear perturbations}
\label{sec3}
 
In this section, we present the equations governing the linear perturbations. For convenience, we work in the Newtonian gauge, where the scalarly  perturbed FLRW metric reads
\be
\label{metric_Newtonian}
ds^2 = - (1+2 \Phi) dt^2  + a^2(t) (1-2 \Psi) d\vec {x}^2 \;.
\ee
For each species, the  continuity  and Euler equations  can be derived from, respectively, the time component and the space components of eqs.~\eqref{evol_matterb}--\eqref{evol_matterc}. As obtained in \cite{Gleyzes:2015pma}, 
 they read in Fourier space
\begin{align}
\dot \delta_b  - \frac{k^2}{a^2}v_b  & =  3\dot \Psi   \, , \label{cont_b} \\
 \dot v_b  & = -  \Phi  \, , \label{eul_b} \\
 \dot \delta_c -\frac{k^2}{a^2}v_c    & = 3 ( \Psi + \gammac H \pi )^{\hbox{$\cdot$}} + 
2({1+\alphaDc})({\alphaCc- 3\gammac }) H (\Phi - \dot \pi)-\alphaDc (\dot \Phi-\ddot \pi )  \label{cont_c} \, , \\
\dot v_c +3H\gammac v_c & =- \Phi  - 3H\gammac \pi \, . \label{eul_c}
\end{align}

These equations must be supplemented by the generalized Einstein equations and by the scalar fluctuation equation. We will not write them explicitly here but they can be found in \cite{Gleyzes:2015pma}.
  
\subsection{Quasi-static approximation}

The evolution of perturbations well inside the horizon is most conveniently studied within the quasi-static approximation.  This is justified for spatial  scales that are smaller than the sound horizon of dark energy, or equivalently for wavenumbers  $k\gg aH/c_s$  (see \cite{Sawicki:2015zya} for a detailed discussion and \cite{Lombriser:2015cla} for a recent analytical extension of this approximation). In this regime, one can 
 neglect time derivatives with respect to space derivatives and the  continuity and Euler equations
 (\ref{cont_b})--(\ref{eul_c}) 
 for the baryon and CDM fluids  simplify into
 \begin{align}
\dot \delta_b  - \frac{k^2}{a^2}v_b  & =  0  \, , \label{conteqb}\\
 \dot v_b  & = -  \Phi  \, , \label{Eulerb} \\
\dot \delta_c -\frac{k^2}{a^2}v_c    & = 0   \, ,  \label{conteqc}\\
\dot v_c +3H\gammac v_c & =- \Phi  - 3H\gammac \pi \, .\label{Eulerc}
\end{align}

The equations for the gravitational potentials $\Phi$ and $\Psi$ 
and for the scalar fluctuation $\pi$ also simplify and become  
constraint equations. The gravitational potentials satisfy two Poisson-like equations, given by~\cite{Gleyzes:2015pma}
\begin{align}
- \frac{k^2}{a^2} \Phi  &=  \frac{3}{2} H^2 \Omega_{\rm m}\left\{ \left(1+\alphaT+ \beta_\xi^2 \right) \omega_b \delta_b +  \left[ 1+\alphaT+ \beta_\xi ( \beta_\xi  +  \beta_\gamma) \right] \omega_c \delta_c \right\}  \, , \label{PoissonPhi}\\
 - \frac{k^2}{a^2} \Psi  &=  \frac{3}{2} H^2 \Omega_{\rm m}\left\{ \left(1+ \beta_{\rm B} \beta_\xi \right) \omega_b \delta_b +  \left[ 1+ \beta_{\rm B} ( \beta_\xi  +  \beta_\gamma) \right] \omega_c \delta_c \right\}  \, ,\label{PoissonPsi}
 \end{align}
where we have introduced the parameters  $\omega_I\equiv {\Omega_I}/{\Omega_{\rm m}}$,
\be
\label{parameters}
\begin{split}
\tbeta_{\rm B}  &\equiv \frac{\sqrt{2}   }{ c_s \DDt^{1/2} }\alphaB  \;,  \\ \tbeta_{\xi} & \equiv \frac{\sqrt{2} }{  c_s \DDt^{1/2} }\xi \equiv  \frac{\sqrt{2} }{  c_s \DDt^{1/2} }\left[ \alphaB (1+\alphaT) + \alphaT - \alphaM\right] \;,
\end{split}
\ee
as well as\footnote{The parameter $\beta_\gamma$ generalizes the parameter $\beta$ defined for coupled quintessence in Sec.~5.3.4 of \cite{Ade:2015rim}. In this case, the relation between the two parameters is $\beta_\gamma = - \sqrt{2} \beta$. We thank Valeria Pettorino for a discussion on this issue.}
\be
\tbeta_{\gamma}  \equiv \frac{3  \sqrt{2}   }{ c_s \DDt^{1/2}  } \gamma_c \;. \label{parameters2}
\ee
The scalar fluctuation also satisfies a Poisson-like equation, which reads
\be
\label{pipoisson}
 - \frac{k^2}{a^2} \pi = 3H \Omega_{\rm m}\frac{ \tbeta_\xi \omega_b \delta_b +  (\tbeta_\xi + \tbeta_\gamma)  \omega_c \delta_c }{\sqrt{2} c_s \DDt^{1/2} }  \;.
\ee

 Combining eqs.~(\ref{conteqb})--(\ref{Eulerc}) with eqs.~(\ref{PoissonPhi})--(\ref{PoissonPsi}) and (\ref{pipoisson}) leads  to a system of two second-order equations for the density contrasts,
\begin{align}
\ddot \delta_b + 2 H \dot \delta_b &= \frac32  H^2  \Omega_{\rm m} \left\{ (1+\alphaT+ \tbeta_\xi^2 ) \omega_b \delta_b +  \left[ 1+\alphaT+ \tbeta_\xi ( \tbeta_\xi  +  \tbeta_\gamma) \right] \omega_c \delta_c \right\}  \;, \label{eqmatterb}\\
\ddot \delta_c + (2 + 3 \gammac) H \dot \delta_c &= \frac32  H^2 \Omega_{\rm m}  \left\{  \left[ 1+\alphaT+ \tbeta_\xi ( \tbeta_\xi  +  \tbeta_\gamma) \right] \omega_b \delta_b +  \left[ 1+\alphaT+  ( \tbeta_\xi + \tbeta_\gamma )^2 \right]   \omega_c \delta_c  \right\}   \;.\label{eqmatterc}
\end{align}
Introducing the  bias $b_c$ ($b_b$) between CDM (baryons) and the total matter density contrast $\delta_{\rm m} \equiv \omega_b \delta_b + \omega_c \delta_c$, as
\be
\label{defbias}
\delta_c =  b_c \, {\delta_{\rm m}} \qquad ( \delta_b =  b_b \, {\delta_{\rm m}} ) \;,
\ee
the influence of modified gravity and nonminimal coupling onto the growth of perturbations enters through the combinations
\be
\label{Ups_g}
\Ups_{b} \equiv \alphaT + \beta_\xi (\beta_\xi +  \beta_\gamma \omega_c b_c) \;, \qquad \Ups_{c} \equiv \alphaT + ( \beta_\xi +  \beta_\gamma) ( \beta_\xi + \beta_\gamma \omega_c b_c) \;,
\ee
which vanish for standard gravity (the friction term $\gamma_c$ on the left hand side of eq.~\eqref{eqmatterc} is essentially a background effect and does not affect directly
the energy density perturbations $\delta \rho_{b,c}$).

Modifications of gravity exchanged by $\pi$ are parametrized by $\beta_\xi$ and the nonminimal coupling of dark matter is parametrized by $\beta_\gamma$~\cite{Gleyzes:2015pma}.  This separation of effects is not physical  
and depends on the choice of frame. Indeed, under a generic change of frame (\ref{transf_cd}), one finds, using (\ref{Mtilde})--(\ref{alphatilde}) 
as well as the relations
\be
\label{changegamma}
\tilde \alpha_{\text{D},I}=\frac{\alphaDI-\alphaD}{1+\alphaD}\, , \quad \tilde \alpha_{\text{C},I}=\frac{\alphaCI-\alphaC}{1+\alphaC}\, , 
\ee
 that these two parameters transform as
\be
\label{transbeta}
\begin{split}
\tilde \beta_\xi & = (\beta_\xi + \beta_{ \gamma *}) (1+\alphaD)^{1/2} \;, \\
\tilde \beta_\gamma  &= (\beta_\gamma  - \beta_{ \gamma *} ) (1+\alphaD)^{1/2} \;, 
\end{split}
\ee
where 
\be
\beta_{\gamma *}= \frac{3 \sqrt{2}  }{c_s \alpha^{1/2}} \gamma_* = \frac{\sqrt{2}}{c_s \alpha^{1/2}} \left[ \alphaC+\frac{\dot \alpha_{\text{D}}}{2H(1+\alphaD)} \right]  \,.
\ee
See also App.~\ref{app:dt} for a discussion on the frame dependence of eqs.~\eqref{eqmatterb} and \eqref{eqmatterc} and of the combinations $\Ups_{b,c}$.

The modification of gravity associated with the parameter $\alphaT$ does not  depend on the exchange of $\pi$, see eq.~\eqref{pipoisson} and Refs.~\cite{Gleyzes:2015pma,Perenon:2015sla} (see also \cite{Jimenez:2015bwa} for a recent discussion on local constraints of this effect), and  does not mix with the other two effects under  change of frame.
We note that 
if $\alphaT\ge0$ (which corresponds to a speed of graviton fluctuations $c_T \ge 1$) in the absence of nonminimal coupling, i.e. 
$\beta_\gamma=0$, the combinations \eqref{Ups_g} are always positive, which tends to enhance the growth of structure. 
More generally, for a positive $\alphaT$  the combinations $\Ups_{b}$ and $\Ups_{c}$ can be negative only if $\beta_\xi$ has the opposite sign of $\beta_\gamma$.

Since equations (\ref{eqmatterb})--(\ref{eqmatterc}) are independent of the wavenumber $k$, one can factorize the time dependence from the $k$ dependence of the initial conditions and write the solutions in the form
\be
\label{solMD}
\delta_c (t, \vec k) = \GG_c  (t) \, \delta_{c,0} (\vec k)\;, \qquad 
\delta_b(t,\vec k) = \GG_b (t)  \, \delta_{b,0}(\vec k) \;,
\ee
where $\d_{c,0}$ and $\d_{b,0}$ represent the initial density contrasts for CDM and baryons respectively,
defined at some earlier time in the matter dominated era.
The two functions of time $\GG_c  (t)$ and $\GG_b  (t)$ are the growth factors for CDM and baryons, respectively, assumed to be equal at the initial time, 
$\GG_c  (0) = \GG_b  (0)=1$.

The continuity equation (\ref{conteqc}) then implies that the velocity potential $v_c$ for CDM  is given by

\beq
\label{vc}
v_c (t, \vec k)=\frac{a^2}{k^2}\dot \GG_c  (t)\, \d_{c,0}(\vec k)= \frac{a^2H}{k^2}\, f_c(t) \, \d_{c}  (t,\vec k)\,,
\eeq
where, in the second equality, we have introduced the CDM growth rate
\beq
\quad f_c\equiv\frac{\text{d} \ln \GG_c}{\text{d} \ln a}\,.
\eeq
Similarly, using the continuity equation (\ref{conteqb}), one finds that the velocity potential $v_b$ for baryons  is given by

\beq
\label{vb}
v_b=\frac{a^2 H}{k^2} f(t)\, \d_{b}(t,\vec k) \, , \qquad 
f_b\equiv\frac{\text{d} \ln G_b}{\text{d} \ln a}\,. 
\eeq

\subsection{Link with observations}
 We now examine how the quantities introduced above can be probed by cosmological observations. 
 
A powerful cosmological probe for dark energy is weak lensing, which depends on the so-called scalar Weyl potential, i.e.~the sum of the two gravitational potentials $\Phi$ and $\Psi$. Combining the Poisson-like equations (\ref{PoissonPhi}) and (\ref{PoissonPsi}), one gets the  expression
 \be
\label{LensingPot}
  \Phi + \Psi=-\frac{3a^2H^2}{2k^2}\Omega_{\rm m}\left[ 2+\alphaT+(\beta_\xi+\betaB) \left(\beta_\xi +\beta_\gamma \omega_c  b_c \right) \right]\delta_{\rm m}\, .
  \ee
  In analogy with the combinations (\ref{Ups_g}), it is convenient to define
 \beq
 \label{Ups_l}
\Ups_{\rm lens} \equiv  \alphaT+(\beta_\xi+\betaB) \left(\beta_\xi +\beta_\gamma \omega_c  b_c \right) \;,
\ee
which vanishes when gravity is standard. 

Another way to probe  dark energy is via the observation of galaxy clustering. In particular, redshift-space distortions are sensitive to the growth rate of fluctuations, which is affected by deviations from standard gravity.  Here we extend previous studies and include also the effect of a nonminimal coupling of CDM.

When observing galaxies, one must take into account the fact that what is directly measured is the redshift, and not the distance of the galaxy. In the parallel plane approximation, the correspondence between the so-called redshift space and real space is described by the change of coordinates  (see e.g.~\cite{Bernardeau:2001qr})
\be
\vec s=\vec x+ \hat z   \frac{v_{\text{g},z}}{aH} \, ,
\ee
where $\vec s$ and $\vec x$ denote the spatial coordinates in redshift  and real space respectively and   $v_{\text{g},z}$ is the  line-of-sight component of the galaxy's peculiar velocity.
At linear order, the invariance of the number of galaxies yields the expression for the number density in redshift space in terms of the number density in real space,
\be
\label{RSD}
\delta_{ \text{g},s} =\delta_{\text{g}}-  \frac{1}{aH}\nabla_z v_{\text{g},z}  \,.
\ee

On large scales, the galaxy peculiar velocity $\vec v_\text{g}$ can be related to the CDM and baryon fluid velocities, respectively $\vec v_b$ and $\vec v_c$, by effectively treating galaxies as test particles (see e.g.~\cite{Chan:2012jj}) of baryon and CDM  mass fractions $x_b \equiv M_b/M_\text{g}$ and $x_c \equiv M_c/M_\text{g}$ ($M_\text{g} \equiv M_b+M_c$), respectively. By considering that the large-scale galaxy momentum coincides with the sum of the baryon and CDM fluids momenta in the linear regime, the galaxy peculiar velocity is given as
\be
\vec{v}_\text{g}=x_c\vec{v}_c+x_b \vec{v}_b \;,
\ee
where $\vec{v}_c=\vec\nabla v_c$ and $\vec{v}_b=\vec\nabla v_b$ are the linear velocities satisfying the Euler equations \eqref{Eulerb} and \eqref{Eulerc}. 
Indeed, in the absence of screening  the mass of the CDM component in the galaxy is not conserved and obeys
\be
\dot{M}_c=3H\gamma_c M_c\,, \label{changeMc}
\ee
in agreement with the background evolution  (since $M_c$ scales as $\rho_c a^3$).
Then, the combination of the Euler equations yields 
\be
\frac{d}{dt}\left(M_\text{g}\vec{v}_\text{g}\right)=\frac{d}{dt}\left(M_c\vec{v}_\text{c}\right)+\frac{d}{dt}\left(M_b\vec{v}_\text{b}\right) =\vec{F}_\text{g},
\ee
where 
\be
\vec{F}_\text{g}=- M_\text{g} \vec \nabla \Phi + 3 H \gamma_c M_c \vec \nabla \pi 
\ee
is the neat force exerted on each galaxy. The last term is due to the fifth force on the CDM component.

Using the expression (\ref{vc}) and (\ref{vb}) for the velocity potentials, one thus finds 
\be
v_\text{g}=\frac{a^2H}{k^2}\left(x_c f_c \, \d_{c}\, +x_b f_b\,  \d_{b}\right)\, .
\ee
Substituting the above expression into (\ref{RSD}), and proceeding as in the standard calculation, one finally obtains,  in Fourier space,
\be
\d_{\text{g},s}=\d_\text{g}+\mu^2\left(x_c f_c  \,\d_{c}\,+x_b f_b \, \d_{b}\right),\qquad \mu\equiv \frac{k_z}{k}\,,
\ee
or 

\be
\d_{\text{g},s}=\d_\text{g}+\frac{\mu^2}{b_\text{g}} ({x_c f_c \, b_c\,+x_b f_b \, b_b} )\d_\text{g}\,,
\ee
after introducing the  galaxy bias $b_\text{g}$,  defined by
\be
\d_\text{g}=b_\text{g} \,  \d_{\rm m}\,.
\ee
The galaxy power spectrum in redshift space is thus given by 
\be \label{galps}
P_{\text{g},s} (\vec k)  =\,  {\big( b_{\rm g}^2+\mu^2  {\fg} \big)}^2 P_{\rm m} (k)   \;,
\ee
where we have introduced the effective growth rate of the galaxy distribution as
\be \label{feff}
\fg \equiv {x_c f_c \, b_c\,+x_b f_b \, b_b} \,.
\ee
In the absence of nonminimal coupling of CDM (i.e. for universally coupled baryons and CDM)  the species have the same velocities, i.e.~$f_b\, b_b=f_c\,b_c=f \equiv  \text{d} \ln \delta_{\rm m} /\text{d} \ln a$.

In the following we will assume the same baryon-to-CDM ratio for each galaxy and we will set this to be the background value, i.e.~$x_c = \omega_c$ and $x_b=\omega_b$. However, one could also consider different populations of galaxies with different baryon-to-CDM ratios and study the effects of equivalence principle violations on large scales between these different populations (see e.g.~\cite{Creminelli:2013nua}).

\section{Parametrization}
\label{sec:timepar}

\renewcommand{\captionfont}{\small}
\newcommand\alphaBz{\alpha_{\text{B},0}}
\newcommand\alphaMz{\alpha_{\text{M},0}}
\newcommand\alphaKz{\alpha_{\text{K},0}}
\newcommand\alphaTz{\alpha_{\text{T},0}}
\newcommand\gammacz{\gamma_{c,0}}
\newcommand\pp{p}
\newcommand\bbi{b_{b,\text{in}}}
\newcommand\bci{b_{c,\text{in}}}

\subsection{Time dependence}
\label{timedep}
As already mentioned,  at the background level the dark energy can be defined by simply giving a specific time evolution for the Hubble parameter. 
For simplicity, we assume that  the expansion history corresponds to that  of $w$CDM, so that $H$ is given by
\begin{align}
H^2 (a) = H_0^2 \left[ \Omega_{\rm m,0} a^{-3} +(1-\Omega_{\rm m,0}) a ^{-3(1+\wDE)} \right] \;,  \label{Hparam}
\end{align}
where $\Omega_{\rm m,0}$ is the fraction of matter energy density today, $w$ is a constant parameter and the scale factor $a$ is normalized to unity today. This choice of parametrization for the background is motivated by the fact that observations suggest that the recent cosmology is very close to $\Lambda$CDM, which corresponds to $w=-1$, and deviations from $\Lambda$CDM in the expansion history are usually parametrized in terms of  $\wDE \neq -1$. In the absence of modifications of gravity and nonminimal couplings, i.e.~for $\alphaM = \gammac=0$, $\wDE$ coincides with the equation of state of dark energy, i.e.~$w_{\rm DE}$ in eq.~\eqref{eos}. Another advantage of this parametrization is that the background expansion remains close to the observed one, even when $\alphaM$ or $\gammac$ are switched on and  matter does not scale as $a^{-3}$ (see eqs.~\eqref{Omegab} and \eqref{Omegac}).  
In this way we can assume that the background cosmological parameters are those fitted by a simple $\Lambda$CDM model.
See discussion at the beginning of Sec.~\ref{sec5} and in App.~\ref{app1}.

In the framework of our effective description, gravitational modifications are encoded in the functions $\alphaB$, $\alphaM$ and $\alphaT$, and the non-minimal coupling of CDM is parametrized by $\gammac$.\footnote{In the quasi-static approximation, the parameter $\alphaK$ does not appear in any equation (note that the combination $c_s^2 \DDt$ does not depend on $\alphaK$), while
$\alphaCc$ and $\alphaDc$ only enter through the combination $\gamma_c$ (the combination $c_s^2 \alpha$ does not depend on $\alpha_{{\rm D}, c}$, since $w_c=0$), so that their individual values remain unconstrained in the analysis.}  The time dependence of these parameters is undetermined
in general. 
In order to obtain some quantitative estimates about how much future observations will be able to constrain these parameters, we will focus in the following on a specific functional form for their time dependence. 

For simplicity, we will assume that the functions $\alphaB$, $\alphaM$ and $\alphaT$ share the same time dependence  $\timedep(t)$,
\be
\begin{split}
\label{defparam}
\alphaB(t)&=\alphaBz \, \timedep(t) \, , \\ 
\alphaM(t)&=\alphaMz \, \timedep(t)\, , \\ 
\alphaT(t) &=\alphaTz \, \timedep(t)\, , \\
\end{split}
\ee
where $\timedep$ is normalized to unity today, i.e. $\timedep(t_0) = 1$, and $\alphaBz$, $\alphaMz$ and $\alphaTz$ denote the current values of these parameters, which we wish to constrain.
To be more specific, we will  consider the following time evolution,\footnote{Another possible choice would be
\be
\Gamma (a) \equiv  \frac{1}{\Omega_{\rm m,0}a^{3\wDE} +(1-\Omega_{\rm m,0}) } \;,
\ee
which has the advantage to be directly related to the scale factor $a$. We have checked that this choice leads to constraints similar  to those obtained with the choice
\eqref{timedep_def}. 
}
\be
\label{timedep_def}
\timedep(t) \equiv \frac{1-\Omega_{\rm m} (t)}{1-\Omega_{\rm m,0}} \;,
\ee
where $\Omega_{\rm m}$ is the total nonrelativistic matter fraction introduced in (\ref{Om}) and 
$\Omega_{\rm m,0}$ its present value. Thus, $\Gamma$ vanishes when the unperturbed energy density of  dark energy is negligible, such as 
at high redshift, and one recovers general relativity. The above parametrization is analogous to the one proposed in \cite{Piazza:2013pua,Bellini:2014fua}, up to a normalization factor.

We parametrize the time dependence of $\gammac$ by assuming that the parameter $\beta_\gamma$, defined in eq.~\eqref{parameters2}, is time-independent, so that 
\be
\gamma_c   (t) = \frac{  \tbeta_{\gamma} }{3  \sqrt{2}   }   \,  c_s (t) \DDt^{1/2}  (t) \; , \label{gamma_evol} 
\ee
and the time dependence on the right-hand side can be computed from  eq.~\eqref{ss}. This choice of parametrisation allows to include coupled quintessence \cite{Amendola:2011ie} as a special case, or more generally other cases where the nonminimal coupling of CDM remains active also when $\dot \phi/(H M)$ becomes negligibly small, since one can have $c_s \DDt^{1/2} =0$ while  $\tbeta_{\gamma}\neq0$. Moreover, $c_s (t) \DDt^{1/2}$ vanishes in matter domination, see App.~\ref{app2} for details.
Therefore, when $\Omega_{\rm m} \to 1$, then $\Gamma \to0$ and $\gamma_c \to 0$, which corresponds to the standard matter dominated phase for the background evolution. However, while modifications of gravity switch off in this limit (i.e. $\alphaB, \alphaM, \alphaT \to 0$), the nonminimal  coupling parametrized by $\beta_\gamma$ remains active (see eq.~\eqref{eqmatterc2} and discussion in the next subsection).

Let us briefly discuss the theoretical constraints coming from the stability conditions \cite{Gubitosi:2012hu,Gleyzes:2013ooa,Gleyzes:2015pma}.
As discussed in Sec.~\ref{sec2}, the absence of ghost-like and gradient instabilities in the tensor  fluctuations respectively requires $M^2 > 0$---which will be always assumed here and in the following---and $c_T^2 > 0$. 
Requiring that the second condition is satisfied at all times, eq.~\eqref{cTsq} implies 
\be
\alphaTz > -1  \;.
\ee 
For scalar fluctuations, these two conditions become 
$\alpha \ge 0$ and $c_s^2 \ge 0$, where the expressions for $\alpha$ and $c_s^2$ are respectively given in eqs.~\eqref{alpha_def} and \eqref{ss}. 
In the following we assume that $\alpha \ge 0$ is  satisfied by an appropriate choice of the parameters $\alphaK$, $\alphaB$ and $\alphaDc$ and we will exclude parameters for which the combination $c_s^2 \alpha$ (see eq.~\eqref{eqtheta}) becomes negative before $z=0$ 
(see again App.~\ref{app2} for details).

\subsection{Initial conditions for the perturbations}

We set the initial conditions during matter domination, i.e.~when $\Omega_{\rm m} \simeq 1$,  and thus $\timedep \simeq 0$. In this limit $\alphaM \simeq 0$ and $\gammac \simeq 0$, so that, according to eqs.~\eqref{cons_1}--\eqref{cons_2}, both CDM and baryons  behave as conserved species at the background level. Moreover, $\alphaT \simeq 0 $  and eqs.~\eqref{parameters}--\eqref{parameters2} respectively imply that $\betaB \simeq 0$ and $\beta_\xi \simeq 0$. Therefore, deep in matter domination eqs.~\eqref{eqmatterb} and \eqref{eqmatterc} simplify to
\begin{align}
\ddot \delta_b + 2 H \dot   \delta_b &\simeq \frac32  H^2 \left[ \omega_b \delta_b +  \omega_c \delta_c \right]  \;, \label{eqmatterb2} \\
 \ddot \delta_c  + 2 H   \dot \delta_c  &\simeq \frac32   H^2  \left[  \omega_b \delta_b +  \left( 1+ \tbeta_\gamma^2 \right)   \omega_c \delta_c  \right]   \;,\label{eqmatterc2}
\end{align}
where $\omega_{b,c}$ are constant. 

This linear system can easily be solved by diagonalizing it. 
One can find solutions written as 
\be
\d_b=\bbi \, \d_{\rm m}\,, \qquad \d_c=\bci\,  \d_{\rm m},
\ee
with constant and scale-independent bias parameters given by 
\be
\label{bias}
\bbi = \frac{1+\beta_\gamma^2 \omega_c - \sqrt{ 4 \beta_\gamma^2 \omega_c^2 + (1-\beta_\gamma^2 \omega_c)^2 }}{2 \beta_\gamma^2 \omega_c \omega_b } \;, \qquad 
\bci = \frac{-1+\beta_\gamma^2 \omega_c + \sqrt{ 4 \beta_\gamma^2 \omega_c^2 + (1-\beta_\gamma^2 \omega_c)^2 }}{2 \beta_\gamma^2 \omega_c^2  } \,.
\ee
The respective growth functions $\GG_c$ and $\GG_b$ are identical, solutions of  the  equation 

\be
\ddot \GG + 2 H \dot \GG - \frac32 H^2 \left(  1 + \beta_\gamma^2 \omega_c^2 \bci   \right) \GG =0 \;. \label{D_evol} 
\ee
As usual, we will consider only the growing mode solution of this equation, $\GG_{+}$. In conclusion, we find that baryons and CDM possess spectra that are initially proportional  and then grow similarly.

Although we use the full expressions from   \eqref{bias} and \eqref{D_evol} in our numerical analysis, it is instructive to consider approximate expressions for small values of $ \tbeta_{\gamma}$.
For small $ \tbeta_{\gamma}$ eq.~\eqref{bias}  yields
\be
\bbi = 1 - \omega_c^2 \tbeta_{\gamma}^2  +\mathcal{O}(\beta_{\gamma}^4)\;, \qquad  \bci = 1 + \omega_c \omega_b   \tbeta_{\gamma}^2  +\mathcal{O}(\beta_{\gamma}^4)\;,
\ee
while the growing solution of eq.~\eqref{D_evol} is of the form
\be
\label{Dplus}
\GG_{+} (a) = a^{1+\frac{3}{5}   \omega^2 _c  \tbeta^2 _{\gamma }} +\mathcal{O}(\beta_{\gamma}^4)\;.
\ee
Thus, for small $ \tbeta_{\gamma}$ the initial conditions in matter domination
are simply given by 
\be
\delta_b (a,k) \simeq (1 - \omega_c^2 \tbeta_{\gamma}^2) \, a^{1+\frac{3}{5}\omega^2 _c  \tbeta^2 _{\gamma }} \, \delta_0(\vec k) \;, \qquad 
\delta_c (a,k) \simeq (1 + \omega_c \omega_b   \tbeta_{\gamma}^2) \, a^{1+\frac{3}{5}\omega^2 _c  \tbeta^2 _{\gamma }} \, \delta_0 (\vec k) \;.
\ee

\subsection{Fiducial models}

For our analysis, we take as fiducial evolution  of the Hubble parameter the function
\beq
\label{H_fid}
\hat{H} (a) = H_0 \sqrt{ \Omega_{\rm m,0} a^{-3} +1-\Omega_{\rm m,0} } \;, \qquad ({\rm Fiducial})
\eeq
which corresponds to the $\Lambda$CDM evolution, i.e.~$\wDE = -1$ in eq.~\eqref{Hparam} and a quantity evaluated on the fiducial model is denoted by a hat. 
The fiducial value for two of the parameters that appear in our analysis is taken to be zero,
\beq
\label{parameters_fid}
\hat{\alpha}_{\text{M},0} = \hat{\alpha}_{\text{T},0} =0 \;, \qquad ({\rm Fiducial})
\eeq
but we consider several options for the parameters $\beta_\gamma$ and $\alphaBz$.
In addition to the simplest case where these parameters are zero, it is also instructive to consider fiducial models where either of these parameters is nonzero.   

We will  distinguish three fiducial models, characterized respectively by the parameters
\begin{description}
\item[I) $\Lambda$CDM:]  $\hat{\alpha}_{\text{B},0}= \hat{\beta}_\gamma=0$,
\item[II) Braiding:] $\hat{\beta}_\gamma=0$, $\hat{\alpha}_{\text{B},0}= - 0.01$,
\item[III) Interacting:] $\hat{\alpha}_{\text{B},0}=0$, $\hat{\beta}_\gamma=-0.03$,
\end{description}
while  the other parameters take the  common values prescribed in (\ref{H_fid}) and (\ref{parameters_fid}).
Case (I) gives the usual  $\Lambda$CDM  for the perturbations. In this case the generalized Einstein equations and the modified continuity and Euler equations reduce to the standard ones. 
Case (II) corresponds to a mixing between the dark energy and gravity kinetic terms at the level of the perturbations. Finally,  in case (III) we allow for  a non vanishing interaction between dark energy and CDM,  which is active for perturbations but does not affect the background because $c_s \alpha^{1/2} =0$, and thus $\gammac = 0$. 
Let us stress that the background evolution is exactly the same for all three fiducial models.

\section{Fisher matrix forecasts}
\label{sec5}

Our constraints will be based on a Fisher matrix analysis applied to the galaxy and weak lensing power spectra \cite{Tegmark:1996bz,Tegmark:1997rp} and to the correlation between the ISW effect in the CMB and the galaxy distribution \cite{Crittenden:1995ak}. In general, the Fisher matrix is defined as
\be
F_{ab } \equiv -{ \bigg\langle \frac{\partial^2 \ln {\cal L} ( \boldsymbol \theta)}{\partial \theta^a \partial \theta^b} \bigg\rangle}_{\hat{\boldsymbol \theta} }\, ,
\ee
where ${\cal L}$ is the likelihood function, $\boldsymbol \theta$ is  a set of parameters.
The expectation values are over realizations. In the fiducial models 
I and 
III 
$\gamma_c$ vanishes when varying along $\beta_\gamma$ (since $c_s\alpha^{1/2}=0$) and thus, since $\beta_\xi =0$ (see eqs.~\eqref{eqmatterb} and \eqref{eqmatterc}), $\beta_\gamma$  only appears quadratically in the perturbation equations. We have checked that observables depend only mildly on $\gamma_c$ for the fiducial 
II. 
Thus, we  choose $\beta_\gamma^2$ rather than $\beta_\gamma$ as the independent variable in the  analysis.
In summary, we have the parameters
\be
\boldsymbol \theta \equiv \{\wDE \, , \, \alphaBz \, , \, \alphaMz \, , \, \alphaTz \, , \, \beta_\gamma^2 \} \;.
\ee

Our goal here is to estimate  the  precision on the above parameters that will be reached by forthcoming spectroscopic and photometric redshift surveys with Euclid-like characteristics \cite{Laureijs:2011gra} (see e.g.~\cite{Zhao:2010dz,Giannantonio:2011ya,Amendola:2011ie} for analogous studies). In particular, we are interested in identifying the degeneracies affecting these parameters and their origin. To simplify this analysis 
we will fix the other background cosmological parameters  to their Planck estimated values: For $w=-1$ these are given by \cite{Ade:2015xua} $h = 0.6731$, $h^2 \Omega_{b,0} = 0.0222 $ and $h^2  \Omega_{c,0} = 0.1197$, while for $w\neq -1$ we  choose the values of $\Omega_{b,0}$ and $\Omega_{c,0}$ such as to maintain the same angular diameter distance as in the $w=-1$ case \cite{Ade:2015xua}.  See details in the App.~\ref{app1}.

\subsection{Galaxy clustering}

The  galaxy power spectrum in redshift space is given by eq.~\eqref{galps}. Including the corrections due to the Alcock-Paczynski effect, 
the observed power spectrum reads \cite{Seo:2003pu} 
\be
P_\text{obs}(z;k,\mu)=  {\cal N} (z)   \left[b_{\rm g}(z) + \fg(z) \mu^2 \right]^2\, P_{\rm m}(z,k) \;,
\ee 
where the normalization factor  ${\cal N} (z)$  is given by 
\be
{\cal N} (z) \equiv \frac{H(z) \hat D_{\rm A}^2(z)}{\hat H(z) D_A^2(z)} \;, \qquad  D_A (z) \equiv \frac{1}{ 1+z} \int_0^z \frac{d \tilde z}{H(\tilde z)} \;,
\ee
and $D_A $ is the  angular diameter distance.
Moreover,  
we assume the bias between galaxies and the total matter distribution,
$b_{\rm g} = \delta_{\rm g} / \delta_{\rm m}$, to be scale independent. 
Its fiducial value has little effects on the constraints; in the following we will assume it to be $\hat b_{\rm g} = \sqrt{1+z}$ \cite{Rassat:2008ja}. It can be taken as a nuisance parameter but we will fix it to its fiducial value, as a consequence of the discussion at the beginning of Sec.~\ref{sec6}. 
Finally,
$\fg$ is given in eq.~\eqref{feff} and $P_{\rm m} (z,k)$ is the total matter power spectrum, given by
\be
P_{\rm m} (z,k) = T_{\rm m}^2 (z)  P_{0} (k)\;, 
\ee
where 
\be
\label{Tm}
T_{\rm m} (z) \equiv \omega_b (z) \, \bbi  \, \GG_b (z) + \omega_c (z) \, \bci  \, \GG_c(z)  \; 
\ee
is the matter transfer function, $P_{0} (k)$ is the initial power spectrum of matter fluctuations, $\d_{\rm m,0}$, during matter domination and $\bbi$, $\bci$ are defined in eq.~\eqref{bias}. As the effects of dark energy and modified gravity intervene at late times, the initial spectrum is independent of the parameters $\boldsymbol \theta$.\footnote{Since modifications of gravity affecting the background evolution
take place only at late time, we are insensitive to the the shift in the matter-radiation equality and to the change in scale of the  power spectrum turnaround described in \cite{Amendola:2011ie}.} 
We have neglected  corrections due to the shot noise in the number of galaxies and the radial smearing due to the redshift uncertainty of the spectroscopic galaxy samples and Doppler shift due to the virialized
motion of galaxies (see e.g.~\cite{Huang:2012mr,Amendola:2012ys}), which become relevant on small scales.

We assume a spectroscopic redshift survey of $15 \, 000$ squared degrees, sliced in eight equally-populated redshift bins (we take the galaxy distribution as given by \cite{Geach:2009tm} with a limiting flux  placed at $4 \times 10^{-16} \, \text{erg} \, \text{s}^{-1} \, \text{cm}^{-2}$) between $z=0.5$ and $z=2.1$.
The 
corresponding
Fisher matrix is given by \cite{Tegmark:1997rp}
\be
F^{\rm LSS}_{ab}(z)= \sum_{\rm bins} \, \frac{V}{2{(2\pi)}^3}\int_{k_{\rm min}}^{k_{\rm max}} 2\pi  k^2 dk \int_{-1}^{1} d\mu\,   \frac{\partial  \ln {P_\text{obs} (z;k, \mu)}} {\partial \theta^a} \frac{\partial  \ln {P_\text{obs} (z;k, \mu)}} {\partial \theta^b}\, ,
\label{Fisher_galaxy}
\ee
where $V$, $k_{\rm min}$ and $k_{\rm max}$ are, respectively, the comoving volume and the minimum and maximum wavenumbers of the  bin. 
In this formula we have neglected the intrinsic statistical error associated with  the white shot noise from the Poisson sampling of the density field \cite{Feldman:1993ky}. However, to be conservative, we choose the maximum wavenumber $k_{\rm max}$ such  that the galaxy power spectrum  dominates over the shot noise and we are well within the linear regime. 
More specifically, for each redshift bin we take  $k_{\rm max}$ as the minimum between $\pi/(2  R)$,  where $R$ is chosen such that the r.m.s.~linear density fluctuation of the matter field in a sphere with radius $R$ is 0.5, and the value of $k$ such that $\bar n_i P_g(k) =1$, where $\bar n_i$ is the number density of galaxies inside the bin. 
We have checked that  these values of $k_{\rm max}$ are always smaller than $H/(\sigma_g (1+z))$, with $\sigma_g = 400\, \text{km} \, \text{s}^{-1}$, i.e.~the scale where the peculiar velocity of galaxies due to their virialized motion becomes important. 
For the minimum wavenumber, we assume $k_{\rm min}=10^{-3} h$ Mpc$^{-1}$. 

Since we work in the quasi-static limit and $P_{0} (k)$ is unaffected by the parameters $\boldsymbol \theta$, the effects of modifications of gravity and nonminimal couplings are scale-independent. Thus,  
the integration over $k$ in eq.~\eqref{Fisher_galaxy} simply gives an overall normalisation to the Fisher matrix.

\subsection{Weak lensing}

For  weak lensing, we consider lensing tomography \cite{Hu:1999ek}. The angular cross-correlation spectra of the lensing cosmic shear for a set of galaxy redshift distributions $n_i(z)$ is given by
\be \label{lenspowspec}
C^{\rm WL}_{ij} (\ell)=  \frac{\ell}{4} \int_0^{\infty} \frac{dz}{H(z)}\, \frac{W_i(z) W_j(z) }{ \chi^3(z)}\, k_\ell^3 (z) P_{\Phi + \Psi} [ z, k_\ell(z)] \;, 
\ee
where $\chi(z) \equiv \int_0^z dz/H (z)$ is the  comoving distance and the lensing efficiency in each bin is given by
\be
W_i (z) \equiv \chi(z) \int_z^{\infty} d \tilde z \, n_i(\tilde z)  \frac{\chi(\tilde z) - \chi(z)}{\chi(\tilde z) } \;,
\ee
with each galaxy distribution normalized to unity, $\int_0^\infty dz \, n_i(z) = 1$.
Moreover,
$P_{\Phi + \Psi} (k)$ is the  power spectrum of  $\Phi + \Psi$. Using eq.~\eqref{LensingPot}, it is related to the matter power spectrum by
\be
P_{\Phi + \Psi} (k)  =  T_{\Phi+\Psi} ^2 (z,k) P_{0} (k) \;,
\ee
where 
\be
\label{TPhiPsi}
T_{\Phi+\Psi} (z,k) \equiv -\frac{3a^2H^2}{2k^2}\Omega_{\rm m}\left[ 2+\alphaT+(\beta_\xi+\betaB) \left(\beta_\xi +\beta_\gamma \omega_c  b_c \right) \right] T_{\rm m} (z) \;
\ee
is the transfer function for $\Psi + \Phi$.
Finally, we define $k_\ell (z) \equiv \ell /\chi(z)$ as the wavenumber which projects into the angular scale $\ell$.

We assume a photometric survey of $15 \, 000$ squared degrees in the redshift range $0 < z < 2.5$, with a redshift uncertainty $\sigma_z (z) = 0.05 (1+z)$, and a galaxy distribution \cite{Smail:1994sx}
\be
n (z) \propto  z^2 \exp\left[ - \left( \frac{z}{z_{ 0}} \right)^{1.5} \right]  \; ,
\ee
where $z_0 = z_{m}/1.412$ and $z_{m}$ is the median redshift, assumed to be $z_{m}=0.9$ \cite{Amara:2006kp,Amendola:2012ys}.
Then, we  divide the survey into 8 equally populated redshift bins. For each bin $i$, we define the distribution $n_i(z)$ by convolving $n(z)$ with a Gaussian whose dispersion is equal to the photometric redshift uncertainty $\sigma_z(z_i) $, $z_i$ being the center of the $i$th bin (see also \cite{Giannantonio:2011ya,Amendola:2011ie}).

Neglecting the shot noise error due to the intrinsic ellipticity of galaxies, 
the Fisher matrix for the cross-correlation spectra in eq.~\eqref{lenspowspec} is given by  \cite{Hu:1998az,Hu:2003pt}
\be
F^{\rm WL}_{ab}= f_{\rm sky} \sum_{\ell = \ell_{\rm min}}^{\ell_{\rm max}} \frac{2\ell+1}{2} \, \text{Tr} \left\{ \frac{\partial  C^{\rm WL}_{ij} (\ell)} {\partial \theta^a}  \big[C^{\rm WL}_{jk}(\ell) \big]^{-1}\frac{ \partial C^{\rm WL}_{km}(\ell)} {\partial \theta^b} \big[C^{\rm WL}_{mi}(\ell) \big]^{-1} \right\} \, ,
\ee
where we choose $\ell_{\rm min} = 10$ and $\ell_{\rm max} =300$. Assuming Euclid-like characteristics \cite{Laureijs:2011gra} for the galaxy density and intrinsic ellipticity noise, we have checked that the chosen $\ell_{\rm max}$ corresponds to scales where the shot noise is negligible and perturbations are only mildly beyond the linear regime at small redshift.\footnote{Notice that the value of $\ell_{\rm max} $ chosen here is smaller than what is usually assumed in comparable analyses (see e.g.~\cite{Amendola:2012ys} and references therein).}

\subsection{ISW-Galaxy correlation}

As a third probe, we consider the cross-correlation between the  ISW effect of the CMB photons and the galaxy distribution in the photometric survey, which is a valuable probe of dark energy and of its clustering properties in the late-time universe (see e.g.~\cite{Hu:2004yd,Corasaniti:2005pq}). We treat the galaxy survey  as for the weak lensing analysis of the previous section, i.e.~we divide it into 8 bins and, for each bin, we consider the same galaxy distribution.
Following \cite{Ho:2008bz}, 
the projected galaxy overdensity in the bin $i$ is given by
\be
g_i (\hat n) =\int_0^{\infty} dz\; n_i(z) b_{\rm g}(z) \delta_{\rm m}[z , \hat n \chi (z)]\;, 
\ee
while the ISW effect is given by
\be
\frac{\Delta T}{T}^{\rm ISW} (\hat n)=-\int_0^{\infty}  {dz} \frac{\partial}{\partial z} \big(  \Phi+  \Psi \big) [z , \hat n \chi (z)] \; .
\ee
With these definitions, the angular power spectra of the projected galaxy overdensity and of the ISW effect are respectively given by
\begin{align}
C_{ij}^{\rm gal} (\ell) &=\int_0^{\infty} dz\; \frac{H(z)}{\chi^2(z)} n_i(z) n_j(z)  {b_{\rm g}^2(z)} \,  P_{\rm m} [z,k_\ell (z)] \; , \\
C^{\rm ISW} (\ell) &=  \int_0^{\infty} dz\;  \frac{H(z)}{\chi^2(z)} \bigg[{\bigg(  \frac{\partial T_{\Phi+\Psi} }{\partial z}  (z,k)   \bigg)}^2 \!   \, P_0 (k ) \bigg]_{k=k_\ell (z)}  \; .
\end{align}
Analogously, the angular cross-correlation spectrum between the ISW effect and galaxies reads
\be
C^{\text{ISW-gal}}_i (\ell)=  - \int_0^{\infty} dz\;  \frac{H(z)}{\chi^2(z)} n_i(z) b_{\rm g}(z) T_{\rm m}(z) \bigg[ \frac{\partial T_{\Phi+\Psi} }{\partial z}  (z,k)    P_0 (k) \bigg]_{k=k_\ell (z)} \;.
\ee

The Fisher matrix for the ISW-galaxy correlation is given by (see e.g.~\cite{Douspis:2008xv,Majerotto:2015bra})
\be
F^{\text{ ISW-gal}}_{ab}= f_{\rm sky} \sum_{\ell = \ell_{\rm min}}^{\ell_{\rm max}} (2\ell+1) \, \frac{\partial  C_{ j}^{\text{ISW-gal}} (\ell) } {\partial \theta^a}  \big[\text{Cov}_{jk}(\ell) \big]^{-1}\frac{ \partial C_{ k}^{\text{ISW-gal}} (\ell) } {\partial \theta^b} \, ,
\ee
where we use $\ell_{\rm min}=10$ and $\ell_{\rm max}=300$ and the covariance matrix is given by
\be
\text{Cov}_{jk}(\ell) =C_{ j}^{\text{ISW-gal}} (\ell) C_{ k}^{\text{ISW-gal}}(\ell) +C^{\rm CMB} (\ell)  C_{ jk}^{\rm gal} (\ell) \; ,
\ee
where $C^{\rm CMB}(\ell)$ is the full CMB angular power spectrum.
We have omitted from this expression the CMB noise, which is negligible for CMB experiments such as WMAP and Planck,  and the galaxy shot noise. We have checked that the latter is small up to  the chosen $\ell_{\rm max}$.

\section{Results}
\label{sec6}

In this section we present the results of the Fisher matrix analysis and the associated degeneracies between parameters.  We start by discussing the  
effects of nonstandard gravity on the evolution of homogeneous quantities. As shown below, they are important to understand the effects on perturbations.

\subsection{Background}

\begin{figure}[h]
\begin{center}
{\includegraphics[scale=0.5]{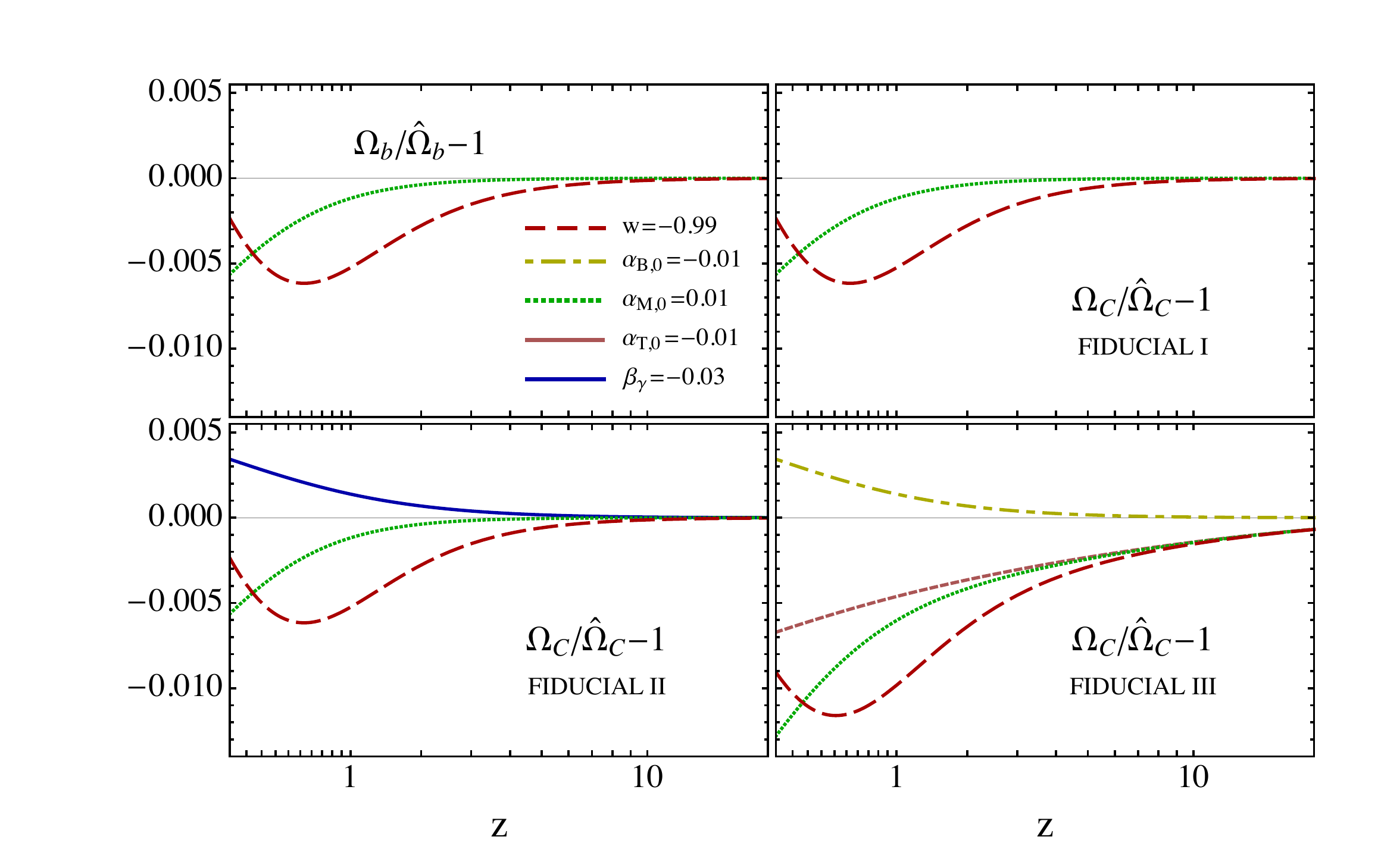}}
\caption{ 
Relative change of the baryon and CDM density fractions,  with respect to their fiducial values, as a function of the redshift $z$, depending on the values of the parameters $w$, $\alpha_{\text{B},0}$, $\alpha_{\text{M},0}$, $\alpha_{\text{T},0}$ and $\beta_\gamma$.}
\label{fig:back}
\end{center}
\end{figure} 

Before presenting the results of the Fisher matrix analysis, we discuss how the background evolution is modified 
when one goes slightly away from any  of the fiducial models by modifying one of the parameters. The results are summarized  in Fig.~\ref{fig:back}, where we have plotted the evolution of the difference between $\Omega_{b,c}$ and their respective fiducial value. 

As is clear from \eqref{Omegab}, the parameter $\Omega_b$ is only affected by a change of the background history embodied by $H(z)$ or by a variation of the effective Planck mass $M$. It is thus only sensitive to a change of the parameters $w$ or $\alphaM$. In the former case, the evolution of $\rho_b$, and thus $\Omega_b$, is modified because $H$ is changed. In the latter case, the evolution of $\rho_b$ does not change but that of $\Omega_b$ does. These changes are independent of the other parameters and one does not need to distinguish between the three fiducial models. 

For $\Omega_c$, the situation is exactly the same as $\Omega_b$ when $w$ or $\alphaM$ are changed, provided there is no coupling between dark energy and CDM, i.e.~$\gamma_c=0$. This is apparent in the boxes corresponding to the fiducial models I and II, for which $\beta_\gamma=0$. By contrast, if we start from the fiducial model III, where $\beta_\gamma\neq0$, and modify either  $w$ or $\alphaM$, then the deviation of $\Omega_c$ with respect to its fiducial value is amplified due to the coupling $\gamma_c$ generated by a nonzero $c_s\alpha^{1/2}$ combined with a nonzero $\beta_\gamma$. For the same reason, i.e.~$\gamma_c\neq0$, we observe a deviation of $\Omega_c$ when $\alphaTz$ or $\alphaBz$ are switched on, in contrast with the other fiducial models. This also explains why one sees a deviation from the fiducial model II when $\beta_\gamma$ is switched on.

The modifications of the background quantities discussed above  affect the observables both indirectly, through their effect on the evolution of perturbations, and directly, because the observables explicitly depend on $H$ and $\Omega_{\rm m}$ (see for instance  eq.~\eqref{TPhiPsi}).
Therefore, a qualitative analysis of the effect of the parameters  $\boldsymbol \theta$ on the observables is rather complex and must take into account both the background evolution and the quantities $\Ups_{b,c}$ and $\Ups_{\rm lens}$. This is why we resort to a Fisher matrix analysis, which allows us to quantify the combined effects on the observables.

\subsection{Forecasts}

\renewcommand{\arraystretch}{1.4}
\begin{table}[t]
\small
\begin{center}
\begin{adjustbox}{max width=\textwidth}
\begin{tabular}{|l|l||c|c|c|c|c|c|c|cl}
  \hline
Fid. & Obs. & $ 10^3 \times \sigma (1+w)$  & $ 10^3 \times \sigma( \alphaBz) $ & $ 10^3 \times \sigma( \alphaMz )$  &   $10^3 \times \sigma(\alphaTz)$  & $10^4 \times \sigma (\beta_\gamma^2)$ \\
 \hline \hline
 I &  GC &  $   7.0 $  & $ 18.6   $ & $   24.5 $  & --  & $1.4 $ \\
  \cline{2-7}
& WL &  $   1.6 $  & $ 4.3   $ & $   42.1 $  & --  &  $5.7  $ \\
  \cline{2-7} 
 & ISW-g &  $   15.5 $  & $ 4.4   $ & $   20.2 $  & --  & $31.3$ \\
  \cline{2-7}
&Comb &  $   1.6 $  & $ 3.0   $ & $   14.6 $  & --  & $1.35$ \\
  \hline
   II &  GC &  $   7.2 $  & $ 18.6   $ & $   33.8 $  & $ 24.4$  & $2.7$ \\
  \cline{2-7}
& WL &  $   1.4 $  & $  4.4  $ & $   67.4 $  & $ 98.9 $  & $ 6.4$ \\
  \cline{2-7} 
 & ISW-g &  $  5.0 $  & $ 4.2   $ & $  24.5 $  & $ 43.2$  &  $56.0$ \\
  \cline{2-7}
&Comb &  $   1.3 $  & $ 3.0   $ & $   19.0 $  & $ 20.8 $  & $2.5$  \\
\hline
   III &  GC &  $   0.22 $  & $  0.40  $ & $   0.22 $  & $ 0.22$  & $1.4$ \\
  \cline{2-7}
& WL &  $   0.17 $  & $ 2.12   $ & $   0.18 $  & $ 0.18$  &  $5.7$\\
  \cline{2-7} 
 & ISW-g &  $   0.88 $  & $ 2.78   $ & $   0.88 $  & $ 0.87$  & $31.3$   \\
  \cline{2-7}
&Comb &  $   0.13 $  & $ 0.39   $ & $   0.14 $  & $ 0.14$  & $1.4$ \\
  \hline
  \end{tabular}
\end{adjustbox}
\end{center}
\caption{ $68\%$ confidence level (CL) 
errors on each individual parameter, assuming that the others take their fiducial values, for each fiducial model and observable: galaxy clustering (GC), weak lensing (WL), ISW-galaxy correlation (ISW-g) and the combination of the three (Comb).\footnotemark   \ The parameter $\alphaTz$ is unconstrained in fiducial model I, see explanation in Sec.~\ref{sec6I}. }
\label{taberrors}
\end{table}

\footnotetext{Our constraints on $\beta_\gamma^2$ are in qualitative agreement with those obtained for coupled quintessence in  \cite{Amendola:2011ie}, taking into account that the parameter $\beta^2$ defined in this reference is related to ours by $\beta_\gamma^2 = 2 \beta^2$. \vspace{0.5mm}}

\begin{figure}[h!]
\begin{center}
{\includegraphics[scale=0.4]{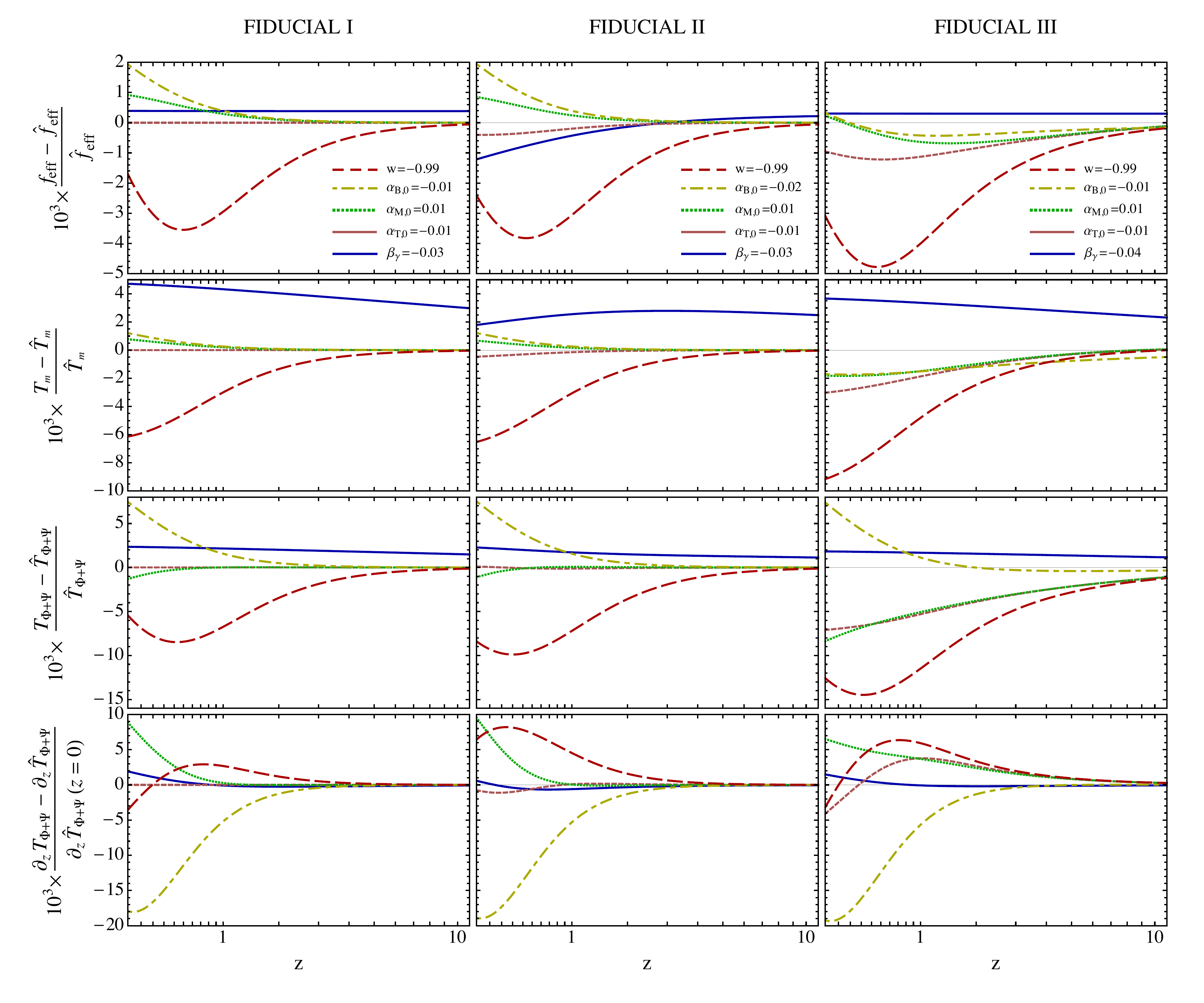}}
\caption{ Modifications of the evolution of perturbations from their fiducial values, as a function of redshift, for the different parameters $w$, $\alphaBz$, $\alphaMz$, $\alphaTz$ and $\beta_\gamma$. From top to bottom,  relative variation of the effective growth factor $\fg$, eq.~\eqref{feff}, the matter transfer function $T_{\rm m}$, eq.~\eqref{Tm}, the Weyl potential transfer function $T_{\Phi +\Psi}$, eq.~\eqref{TPhiPsi} and its derivative with respect to  redshift, $\partial_z T_{\Phi +\Psi}$, for the three different fiducial models (respectively I, II and III, from left to right). As $\partial_z T_{\Phi +\Psi}$ vanishes in matter domination, we have normalized it to its value at $z=0$ instead of its value as a function of the redshift.   }
\label{fig:pert}
\end{center}
\end{figure}

Let us now discuss the results of the Fisher matrix analysis. The unmarginalized   errors on the parameters are summarized in Tab.~\ref{taberrors} while the two-dimensional  contours are presented in Figs.~\ref{fig:FM1}, \ref{fig:FM3} and \ref{fig:FM2}. Red dotted, green dashed and yellow solid lines respectively correspond to galaxy clustering, weak lensing and ISW-galaxy observables. The combination of the three observables, given by summing the three Fisher matrices, is plotted in thick solid black line. The shaded blue regions in the plots correspond to instability regions, where $c_s^2 \alpha < 0$.\footnote{Here we conservatively exclude the instability region from the allowed parameter space. A more refined 
treatment would require multiplying the likelihood function by a theoretical prior that excludes the  forbidden region, which is impossible to achieve with a Fisher matrix analysis (our priors cannot be represented with an invertible matrix).   }

For each observable, the Fisher matrix  including all the parameters is ill-conditioned and cannot be inverted. This means that the observables do not have the constraining power   to resolve the degeneracies (see e.g. \cite{Vallisneri:2007ev}).  
Thus, when plotting the two-dimensional contours we do not marginalise over the other parameters but we fix them to their fiducial values.

As shown in Tab.~\ref{taberrors}, the forecasted constraints from the three probes for the same fiducial model are comparable, within an order of magnitude. This reflects the comparable effects on the observables, shown in Fig.~\ref{fig:pert}, given our choice of $k_{\rm max}$ and $\ell_{\rm max}$  for the spectroscopic and photometric surveys, respectively, which translates into a comparable number of modes for the three probes. More precisely, the effects of gravity modifications and nonminimal couplings is slightly larger on the  lensing potential and ISW effect  but this is compensated by a larger number of modes in the spectroscopic survey. 

Specifically, for this survey the number of modes is roughly given by $N_{\rm modes} \sim N_{\rm bins} \times V \times (4 \pi/3) (k_{\rm max} / 2 \pi)^3 $, where $N_{\rm bins} =8$ is the number of bins and $V  $ is the (average) comoving volume of the bins. Assuming $k_{\rm max} = 0.1 \, h \text{Mpc}^{-1}$, this yields $N_{\rm modes} \sim  10^6  $. For the photometric survey we have $N_{\rm modes} \sim N_{\rm bins}   \times f_{\rm sky} \times \ell_{\rm max}^2  \sim 3 \times 10^5 $. As a rule of thumb, the relative effects of $\alphaBz$, $\alphaMz$ and $\alphaTz$ on the three observables are typically of the order of $ {\cal O} (0.1)$ at redshift $z \sim 1$, see Fig.~\ref{fig:pert}. Thus, one expects to be able to constrain these parameters at the  level of $ {\cal O} (0.1)^{-1} \times N_{\rm modes}^{-1/2}$, i.e.~few percents (which is improved by one order of magnitude 
 for fiducial III, where the effects on the observables are larger), if all the other parameters are fixed. 
The ISW-galaxy correlation is limited by cosmic variance but due to the larger sensitivity of $\partial_z T_{\Phi +\Psi}$ to the modifications of gravity, it sometimes provides constraints comparable to those from the other probes.\footnote{We thank Alessandro Manzotti and Scott Dodelson for pointing out a numerical underestimation of the noise in the ISW-galaxy correlation in an earlier version of this paper, corrected here.}
The effect of $\beta_{\gamma}^2$ is typically of the order of a few at redshift $z \sim 1$ and this parameter can be constrained at a level of a  $\text{few} \times 10^{-4}$ for galaxy clustering and weak lensing. Given  the smaller effect on the ISW and the smaller number of modes for the photometric survey, the ISW-galaxy correlation  provides the weakest constraints on this parameter. We also notice that the degeneracy of this parameter with the others is rather small.

\subsubsection{Fiducial I: $\Lambda$CDM}
\label{sec6I}

\renewcommand{\arraystretch}{1.4}
\begin{table}[t]
\small
\begin{center}
\begin{adjustbox}{max width=\textwidth}
\begin{tabular}{|c||c|c|c|c|c|c|c|cl}
  \hline
  Obs. & Fiducial I &  Fiducial II &  Fiducial III    \\
 \hline\hline
 GC &  $   (0.012 , - 0.007, 0.005, 0 ,1)
$  
& $     ( 0.022   , - 0.013, 0.007 ,  0.01, 1 )
$ 
& $  (-0.626 , 0.348  ,-0.629 ,0.64 , 1) 
$  \\
  \hline
 WL &  $   (-0.345 , - 0.115, - 0.007, 0 ,1)  
$  
& $    ( -0.463, - 0.136 , - 0.001 ,  0.004, 1) 
$ 
& $  (1, -0.074 , 0.910 , -0.914 ,-0.293 ) 
$  \\
  \hline 
  ISW-g &  $   (0.053 , 0.7, 0.154, 0 , 1) 
$  
& $    ( 0.856,  1,  0.117 , 0.063 , -0.609) 
$ 
& $    (-0.997 ,- 0.138 , -0.989 , 1 , -0.068)  
$  \\
  \hline
Comb. &  $   (-0.008 , - 0.012, 0.005, 0 , 1) 
$  
& $    (-0.055 , -  0.034 , 0.006 , 0.009 , 1)
$ 
& $    (1,- 0.285 , 0.953 ,  -0.964 , -0.867 )  
$  \\
  \hline
  \end{tabular}
\end{adjustbox}
\end{center}
\caption{First  eigenvector of the Fisher matrices, for the basis $\{ w, \alphaBz, \alphaMz, \alphaTz, \beta_\gamma^2\}$, with the maximum eigenvalue, corresponding to the combinations of parameters that are maximally constrained by experiments. The coefficients are normalized by the maximum component and rounded to three significant digits. }
\label{tabfiducial}
\end{table}

\begin{figure}[h!]
\begin{center}
{\includegraphics[scale=0.475]{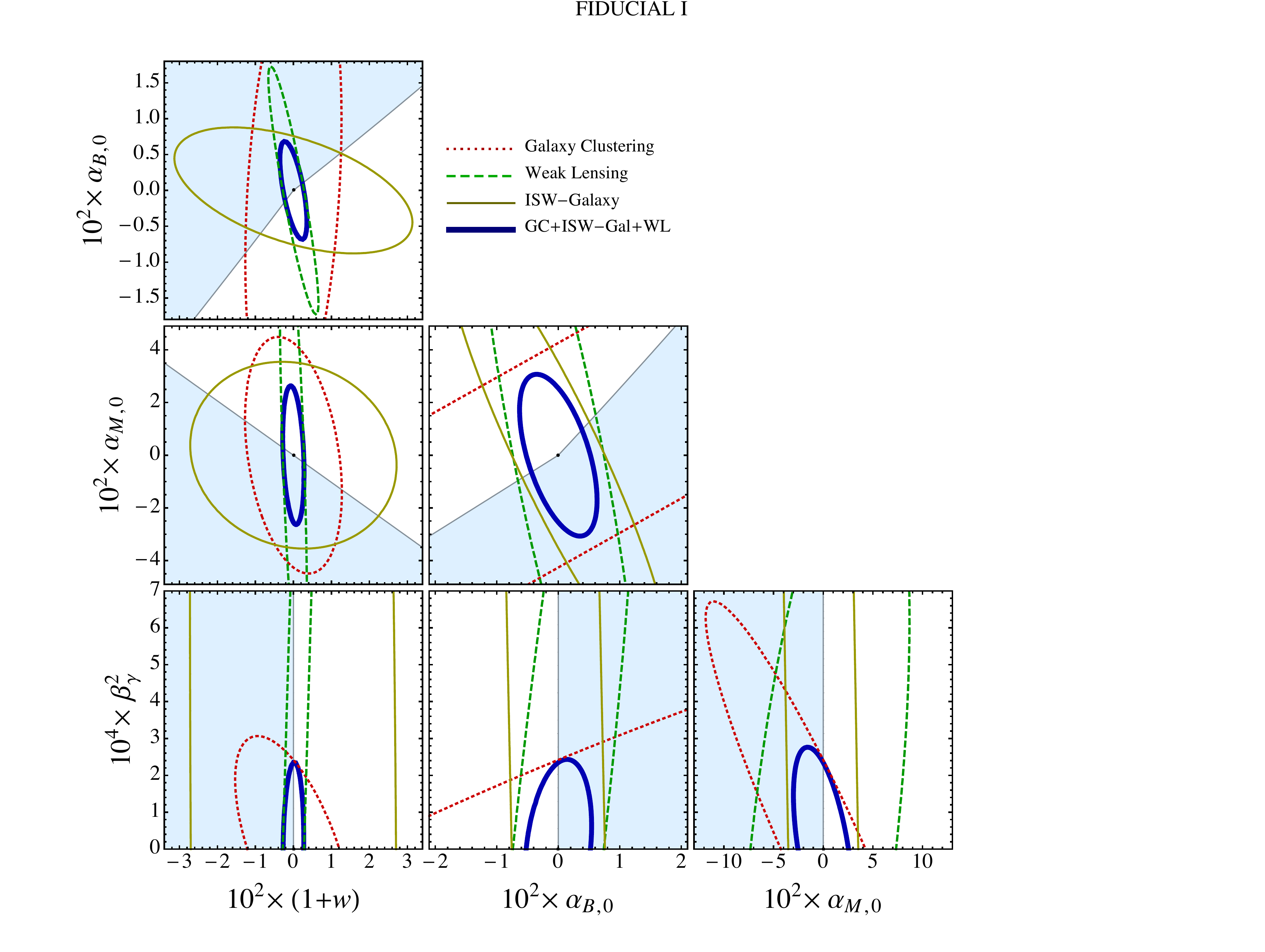}}
\caption{ Two-dimensional $68\%$ CL  contours for the fiducial model I ($\Lambda$CDM model), obtained by fixing all the other parameters to their fiducial values. The parameter $\alphaTz$ is absent, as it is unconstrained on this fiducial model. Shaded blue regions correspond to theoretically forbidden parameter space where $c_s^2 \alpha<0$. Note that the axis range is different for different parameter planes. }
\label{fig:FM1}
\end{center}
\end{figure}

Let us study the constraining power of the observables around a $\Lambda$CDM model.  The 
errors are reported in Tab.~\ref{taberrors}
 and the $68\%$ CL contours are shown in Fig.~\ref{fig:FM1}. 
 In Tab.~\ref{tabfiducial} we report, for each Fisher matrix, the eigenvector associated to the maximal eigenvalue (called here maximal eigenvector), which provides the direction maximally constrained in parameter space, i.e.~the one that minimizes the degeneracy between parameters.

At first view, the parameter $\alphaTz$ seems to contribute to the growth of perturbations  through the combinations $\Ups_b$ and $\Ups_c$, defined in
 (\ref{Ups_g}), and  to the lensing potential through the combination $\Ups_{\rm lens}$, given in 
 (\ref{Ups_l}). However, it turns out  that these combinations in fact do not depend on $\alphaT$ for this choice of fiducial model.

  More precisely, when $w=-1$ and $\beta_\gamma=0$, one finds that  
\be
\Ups_{b,c}=\alphaT+\beta_\xi^2=\alphaT+\frac{2\xi^2}{c_s^2\alpha}, \qquad c_s^2\alpha=-2(1+\alphaB)\xi+3\Omega_{\rm m} \alphaB-3\frac{\dot\alphaB}{H}\,.
\ee
When one goes away from the fiducial model by switching on the parameter $\alphaT$, while all the other parameters keep their fiducial value, one gets $\beta_\xi^2=-\alphaT$ so that the dependence on $\alphaT$ vanishes in $\Ups_{b,c}$. It is immediate to check that $\alphaT$ disappears in $\Ups_{\rm lens}$ for the same reason. 
Thus, the parameter $\alphaTz$ cannot be constrained by a Fisher matrix analysis for this choice of fiducial and will be dropped from the analysis in this subsection. Correspondingly, 
the component in the $\alphaTz$ direction of the maximal eigenvectors vanishes, see Tab.~\ref{tabfiducial}.

Let us now examine the situation when $\alphaB$ is switched on while all the other parameters take their fiducial value.  The $\Ups$ combinations  
are then given by 
\beq
\Ups_{b,c}=\frac{2\alphaB^2}{c_s^2\alpha}\,,\qquad      \Ups_{\rm lens}=\frac{4\alphaB^2}{c_s^2\alpha}\,,
\eeq
with
\beq
c_s^2\alpha=-(2+3\Omega_{\rm m})\alphaB-2\alphaB^2\,.
\eeq
For small values of $\alphaB$, 
we thus find
\beq
\label{f1_aB}
\Ups_{b,c}\simeq -\frac{2}{2+3\Omega_{\rm m}} \alphaB\,,\qquad \Ups_{\rm lens}\simeq -\frac{4}{2+3\Omega_{\rm m}} \alphaB\,.
\eeq
 Thus, one expects the impact of $\alphaB$ to increase
 as $\Omega_{\rm m}$ diminishes, which is in agreement with the results plotted in Fig.~\ref{fig:pert}.

When one changes $\alphaM$ from its fiducial value (the other parameters keeping their fiducial value), one finds
\beq
\label{f1_aM}
\Ups_{b,c}= \Ups_{\rm lens}=\alphaM\,.
\eeq
 As seen in Fig.~\ref{fig:pert}, the effect of $\alphaM$ and $ \alphaB$ on the growth of structures (i.e.~on $\fg$ and $T_{\rm m}$) is roughly the same in magnitude but opposite in sign, which is in agreement with the relations found in (\ref{f1_aB}) and (\ref{f1_aM}). This qualitatively explains the degeneracy observed in the $\alphaBz$--$\alphaMz$ panel of Fig.~\ref{fig:FM1} for galaxy clustering and the corresponding components of the maximal eigenvectors  in Tab.~\ref{tabfiducial}. By contrast, the degeneracy between $\alphaB$ and $\alphaM$ observed for weak lensing does not seem to agree with the values of 
 $\Ups_{\rm lens}$ in (\ref{f1_aB}) and (\ref{f1_aM}). The reason for this discrepancy is that the background is also modified when $\alphaM\neq 0$, as discussed earlier, whereas the background for $\alphaB\neq 0$ is the same as the fiducial one. Since the transfer function $T_{\Phi+\Psi}$ depends not only on the coefficient $\Ups_{\rm lens}$ but also on the background, the degeneracy is more complex. In fact, the background modification also affects the matter growth but more modestly than for weak lensing. 
 
To conclude, let us note that a large region of the observationally constrained parameter space is forbidden by the stability requirements, i.e.~$c_s^2 \alpha>0$.

\subsubsection{Fiducial II: Braiding}

\begin{figure}[h!]
\begin{center}
{\includegraphics[scale=0.39]{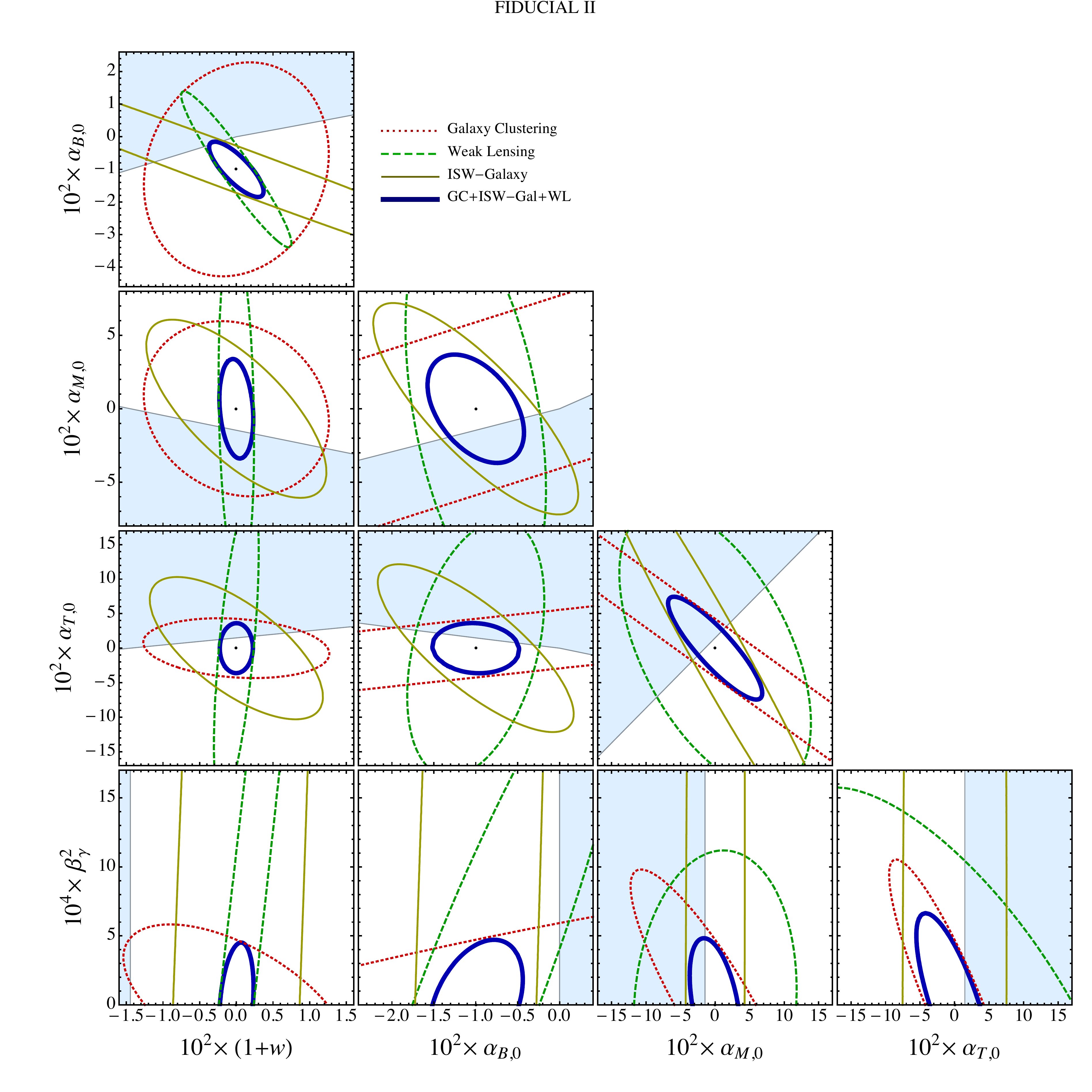}}
\caption{  Two-dimensional $68\%$ CL  contours  for the fiducial model 
II 
(braiding model with $\alphaBz=-0.01$), obtained by fixing all the other parameters to their fiducial values. Shaded blue regions correspond to $c_s^2 \alpha<0$. The axis range is different for different parameter planes.   }
\label{fig:FM3}
\end{center}
\end{figure}

For this fiducial model, we have the value
  $\hat{\alpha}_{\text{B},0} = - 0.01$, where the negative sign is to satisfy the stability conditions. This corresponds to dark energy models where the kinetic term of $\pi$ comes from a mixing with gravity \cite{Creminelli:2006xe,Creminelli:2008wc}, which are sometimes called braiding models \cite{Deffayet:2010qz,Pujolas:2011he}.
The unmarginalized
errors are reported in Tab.~\ref{taberrors} and the $68\%$ CL contours are shown in Fig.~\ref{fig:FM3}. Note that the allowed parameter space is much larger than in the previous fiducial because for $\alphaBz \neq 0$ the null energy condition can be violated without  instabilities \cite{Creminelli:2006xe}.

In this case, $\Ups_{b,c}$ and $\Ups_{\rm lens}$ depend on  $\alphaT$: their partial derivatives with respect to $\alphaT$ on the fiducial model are given by
\be
\frac{\partial \Ups_{b,c}}{\partial \alphaT} =\frac{9 \Omega_{\rm m}^2}{(3 \Omega_{\rm m}+2 + 2 \alphaB )^2} \;, \qquad \frac{\partial \Ups_{\rm lens}}{\partial \alphaT}   =\frac{3 \Omega_{\rm m} (3 \Omega_{\rm m}-2 - 2 \alphaB )}{(3 \Omega_{\rm m}+2 + 2 \alphaB )^2}  \;,
\ee
which confirms that this parameter must be included in the analysis.

For this fiducial, the plane $\alphaBz$--$\alphaTz$ in Fig.~\ref{fig:FM3} has the same background evolution as $\Lambda$CDM. Therefore, all the effects are controlled by $\Upsilon_{b,c}$ and $\Upsilon_{\rm lens}$, so that the degeneracies can in principle be understood analytically from their expressions in terms of $\alphaBz$ and $\alphaTz$.  For instance, for small $\alphaBz$ and $\alphaTz$ one finds
\be
\Upsilon_{b,c}  \simeq \frac{3   \alpha _{\text{B},0} \left(\Omega _{\rm m}-1\right) \left(2 \alpha _{\text{B},0}+\left(2-3 \Omega _{\rm m}\right) \alpha _{\text{T},0}\right)}{\alpha _{\text{B},0} \left(6 \Omega _{\rm m}+4\right)+4 \alpha _{\text{T},0}}  \simeq  (1-\Omega _{\text{m}}) \left(0.54 \alpha _{\text{T},0}-0.6 \Delta \alpha _{\text{B},0}\right) \;, 
\ee
where in the last equality we have expanded at linear order for small $1-\Omega_{\rm m}$ and used $\alphaBz = -0.01 + \Delta \alphaBz$. This explains the degeneracy between $\Delta \alphaBz$ and $\alphaTz$ observed in the growth. By the same procedure we find $\Ups_{\rm lens} \simeq (1-\Omega _{\text{m}}) \left(0.18 \alpha _{\text{T},0}-1.2 \Delta \alpha _{\text{B},0}\right) $, which explains why $\Delta \alpha _{\text{B},0}$ is more constrained than $ \alphaTz$ by lensing observations.

Similarly to fiducial I, the effect of changing $\alphaBz$ and $\alphaMz$ on the growth of structures is roughly the same in magnitude and opposite in sign. This effect can be qualitatively understood by expanding $\Ups_{b,c}$ for small $\Delta \alphaBz$ and $\alphaMz$, analogously to what was done in Sec. 6.2.1. This degeneracy cannot be seen for the lensing, because the modifications of the background also play a role.

\subsubsection{Fiducial III: Interacting}

\begin{figure}[h!]
\begin{center}
{\includegraphics[scale=0.39]{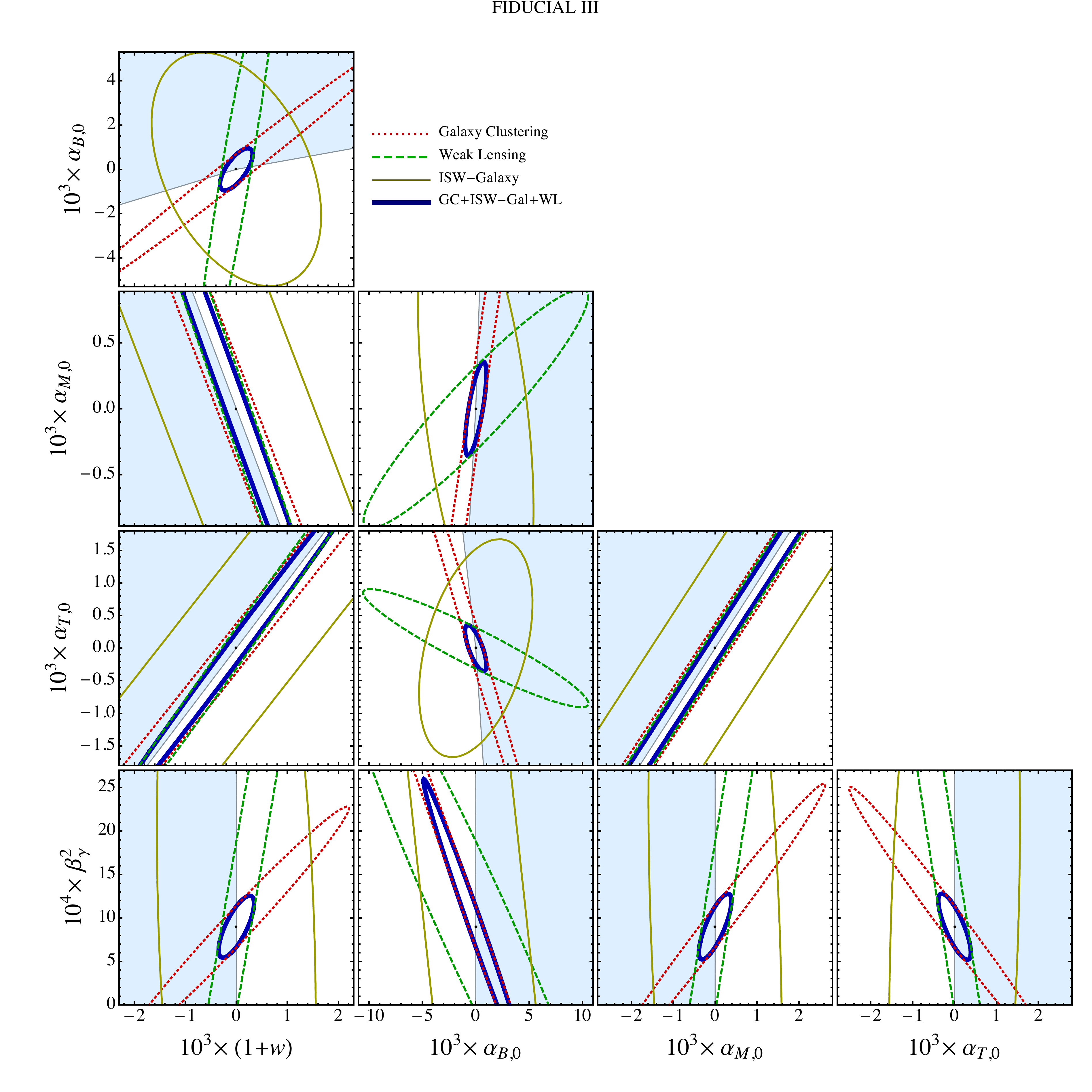}}
\caption{  Two-dimensional $68\%$ CL  contours  for the fiducial model 
III 
(interacting model with $\hat{\beta}_\gamma = -0.03$), obtained by fixing all the other parameters to their fiducial values. Shaded blue regions correspond to $c_s^2 \alpha<0$. The axis range is different for different parameter planes.  }
\label{fig:FM2}
\end{center}
\end{figure}

In this  model we have a nonzero fiducial value for the parameter $\beta_\gamma$ ($\hat{\beta}_\gamma = - 0.03$), which implies an active coupling between CDM and dark energy. The unmarginalized
errors are reported in Tab.~\ref{taberrors} and the $68\%$ CL contours are shown in Fig.~\ref{fig:FM2}. Notice that the constraints for this fiducial model are generally stronger than those for models I and II (see below). As one can verify in Fig.~\ref{fig:pert}, this is due to the enhancement of the effects on the observables, caused by the nonminimal coupling.

In this case, $\alphaTz$ must be included in the analysis, because 
  $\Ups_{b,c}$ and $\Ups_{\rm lens}$ depend on  $\alphaTz$  through the term $\beta_\xi \beta_\gamma$.
Indeed, let us  examine the case when $\alphaTz$ and $\alphaMz$ are switched on
while $w=-1$ and $\alphaBz=0$.  Using $c_s \alpha^{1/2} =  \sqrt{2(\alphaM- \alphaT)}=  - \sqrt{2} \beta_\xi $  (we assume $\alphaM>\alphaT$ to satisfy the stability condition) one finds
\beq
\label{upsglFid2}
  \Ups_{b}= \alphaM - \sqrt{\alphaM- \alphaT} \, \beta_\gamma \omega_c b_c \,, \qquad  \Ups_{c}=   \alphaM -  \sqrt{\alphaM- \alphaT} \, \beta_\gamma (1+\omega_c b_c) + \beta_\gamma^2 \,
\eeq
and
\be
    \Ups_{\rm lens}= \alphaM -  \sqrt{\alphaM- \alphaT} \, \beta_\gamma  b_c \omega_c \,.
\ee
However, the degeneracies  observed in Fig.~\ref{fig:FM2}, for example in the plane $\alphaMz$--$\alphaTz$, cannot be understood directly from the above expressions because, as we saw in Fig.~\ref{fig:back},  the background is modified, not only when $\alphaM$ (or $w$) is changed but also when $\alphaT$ is changed.

Another notable degeneracy appearing in Fig.~\ref{fig:FM2} is between $w$ and the parameters $-\alphaTz$ or $\alphaMz$. This can be partially understood  from the fact that $w$ appears in the combination  
\be
\label{wxi}
c_s^2 \alpha \simeq 3 (1+w) (1-\Omega_{\rm m})- 2 (\alphaM - \alphaT)  = 3 (1-\Omega_{\rm m}) \left(1+w  - \alphaMz + \alphaTz \right)  \;,
\ee
where we have used $\eta \simeq - w (1-\Omega_{\rm m})$ in  eq.~\eqref{csalpha} for the first equality and $\Omega_{\rm m,0} \simeq 1/3$ in the last one.
However, 
background effects play an important role as well.

The term $\beta_\xi\beta_\gamma$ in eqs.~\eqref{Ups_g} and \eqref{Ups_l} translates here as $- \sqrt{\alphaM-\alphaT}\, \beta_\gamma$, see eq.~\eqref{upsglFid2}. This term encodes the new effects that arise when both modifications of gravity and nonminimal couplings are considered, as  emphasized in \cite{Gleyzes:2015pma}. 
These effects explain the qualitative difference, in the size and shape, between the contours of fiducial 
III 
(Fig.~\ref{fig:FM2}) and those of the other two fiducial models. Not only are the constraints tighter by an order of magnitude in this case, but also the maximal eigenvectors of the Fisher matrices point in different directions, see Tab.~\ref{tabfiducial}.

\section{Summary and conclusions}
\label{conclusion}

In this paper, we have investigated the consequences of both modifying gravity and allowing a coupling between CDM and dark energy. If the 
propagation speed of dark energy is not too small, one can rely on the quasi-static approximation because the small scale fluctuations of dark energy have the time to relax to the quasi-static regime \cite{Sawicki:2015zya}. In this case, the  parameters describing  deviations from $\Lambda$CDM, which are usually four for Horndeski-like theories \cite{Gleyzes:2013ooa,Bloomfield:2013efa,Bellini:2014fua}, reduce to three: $\alphaB$, $\alphaM$ and $\alphaT$ \cite{Gleyzes:2015pma}. Moreover, the coupling of a fluid of CDM particles conformally and disformally coupled to dark energy,  can be described by a single parameter  $\gamma_c$, see eq.~\eqref{gammac}.

The dynamics of matter perturbations also simplifies. In particular, as discussed in Sec.~\ref{sec3}, it is described by a  system of two coupled equations, eqs.~\eqref{eqmatterb} and \eqref{eqmatterc}, respectively for baryons and CDM. In these equations, the four parameters above enter in three combinations (see eq.~\eqref{Ups_g}):  $\alphaT$,  $\beta_\xi$ (a combination of $\alphaB$, $\alphaM$ and $\alphaT$) and  $\beta_\gamma$, the latter describing the nonminimal coupling of CDM perturbations.
 As explained in more details in Sec.~\ref{sec3}, these distinctions are frame-dependent, as one can verify using the relations \eqref{alphatilde} (see also \cite{Gleyzes:2015pma} for more details).

The growth of fluctuations is usually described in terms of the growth rate, which  modulates the galaxy power spectrum in redshift space and  can thus be measured with redshift space distortions. We have computed the effective growth rate for galaxies made of baryons and nonminimally coupled CDM, in the presence of modifications of gravity. This is the first general treatment of this kind, to our knowledge.

Deviations from the $\Lambda$CDM model can also affect the propagation of light through their effect on the scalar Weyl potential, i.e.~the sum of the two metric potentials in Newtonian gauge. A fourth parameter, 
 $\beta_{\rm B}$ (proportional to $\alphaB$),
together with the three parameters above, is necessary to fully describe this effect, which can be measured in the weak lensing  and ISW effect (see eq.~\eqref{Ups_l}). 

As discussed in Sec.~\ref{sec:timepar}, the evolution of perturbations depends on the time dependence of the Hubble rate and of the parameters described above. In the present work we have taken the Hubble rate to be the same as in $w$CDM. Moreover, the parameters $\alphaB$, $\alphaM$ and $\alphaT$ grow as $1 -\Omega_{\rm m}$, so that modifications of gravity disappear in matter domination, while the nonminimal coupling remains active at all times, i.e.~$\beta_\gamma=$ constant. We have studied the constraining power  of a future redshift survey with Euclid specifications on the parameters $w$, $\alphaBz$, $\alphaMz$ and $\alphaTz$.

More specifically, in Sec.~\ref{sec5} we computed the Fisher matrix of the galaxy power spectrum, the weak lensing power spectrum as well as the correlation spectrum between the ISW effect and the galaxy distribution. We have considered five parameters, namely
$w$ (describing the background evolution), the current values of $\alphaB$, $\alphaM$ and $\alphaT$, and the constant nonminimal coupling parameter $\beta_\gamma^2$, and assumed three fiducial models: (I) $\Lambda$CDM, 
(II) a braiding model with $\alphaBz = -0.01$  and  (III) an interacting model with $\beta_\gamma = -0.03$.

The unmarginalized $68\%$ CL errors on these parameters are reported in Tab.~\ref{taberrors} in Sec.~\ref{sec6}.
For the current values of $\alphaB$, $\alphaM$ and $\alphaT$, the errors are of the order of  $ 10^{-2}$--$10^{-3}$  for fiducial models I and 
II and an order of magnitude better for the fiducial model III. 
The error on  $\beta_\gamma^2$ is of the order of $10^{-4}$ for all fiducial models.  Given the large number of free parameters and the degeneracies among them, the Fisher matrices cannot be inverted to compute the marginalized contours. Therefore, 
we have shown the two-dimensional $68\%$ CL contours in Figs.~\ref{fig:FM1}, \ref{fig:FM3} and \ref{fig:FM2}---together with the excluded parameter space from stability conditions---by setting  all the other free parameters to their fiducial values. Moreover, we have provided a discussion on the origin of the degeneracies and the constrained directions in parameter space in Tab.~\ref{tabfiducial}.
As shown by the contour plots, all the three observational probes are complementary in breaking degeneracies in parameter space.

This analysis can be generalized in several directions. First, the background cosmological parameters should be included in the analysis as nuisance parameters. In this case, it is important to take as well into account other cosmological data such as the CMB, the baryon acoustic oscillations and the supernovae Type Ia.  Another direction is exploring alternative parametrizations of the background evolution and/or of the time dependence of the parameters $\alphaB$, $\alphaM$, $\alphaT$ and $\beta_\gamma$. For instance, assuming that the $\alpha$'s vanish at early times, as we did, considerably limits the effect of dark energy on certain observables such as the CMB or the matter power spectrum. On the other hand, one could assume other  equally  motivated time dependencies (even different for different parameters), which are expected to lead to larger effects in the observables.
The final goal is to extend this analysis beyond the quasi-static approximation to include larger scales and other species, such as neutrinos and photons. Such a program has been  initiated with the development of the publicly available Boltzmann codes EFTCAMB \cite{Hu:2013twa} (see \cite{Frusciante:2015maa} for a recent application to Horava gravity) and COOP \cite{Huang:2015srv}. In this case, at least one more parameter, $\alphaK$, must be considered in the analysis.  
On the other hand, one may expect that some of the degeneracies found in this paper can be resolved by the scale dependence appearing once the full dark energy dynamics is taken into account. We leave this for future work.

\vspace{5mm}
\noindent
{\bf Acknowledgements}.
We wish to thank Tessa Baker, Phil Bull, Scott Dodelson, Pedro Ferreira, Zhiqi Huang, Alessandro Manzotti, Valeria Pettorino, Federico Piazza, Ignacy Sawicki and Emiliano Sefusatti for stimulating discussions. J.G., M.M.~and F.V.~thank the APC ({\em AstroParticule et Cosmologie}) and PCCP (Paris Center for Cosmological Physics) for kind hospitality. F.V.~acknowledges partial support from the grant ANR-12-BS05-0002 of the French Agence Nationale de la Recherche.

\appendix

\section{Details on the parametrization}

In this appendix we provide some details about the determination of the background parameters in our numerical calculations  and about  the value of the effective functions in our parametrization. 

\subsection{Background quantities}
\label{app1}
Assuming that gravity is standard at recombination, dark energy can only affect the best fit value of the cosmological parameters inferred through the measurement of the comoving distance to last scattering with the CMB spectrum. Thus, we assume that the comoving distance to last scattering is fixed and given by its best fit measurement \cite{Ade:2015xua} and we compute the values of the background cosmological parameters inferred from this observation. Let us discuss how these are determined. When $w=-1$, these are chosen as the base $\Lambda$CDM best fit values of the {\em Planck} TT+lowP parameters \cite{Ade:2015xua}.
When $w\neq -1$, we determine the initial conditions for the background matter components by requiring the comoving distance\footnote{The comoving distance is related to the luminosity distance $D_L$ and the angular-diameter distance $D_A$ by the relations   $D_L(z)=(1+z) \chi(z)$ and $D_A(z)= \chi(z)/(1+z)$.} 
\beq
\chi(z_{\rm in}; \Omega_{\rm m,0}, w)=\int_0^{z_{\rm in}} dz H^{-1}(z;\Omega_{\rm m,0}, w)
\eeq
to be the same as the one of the $\Lambda$CDM model. More precisely, for each value of $w$, we associate the parameter $\Omega_{m,0}(w)$ defined by the relation
\beq
\chi(z_{\rm in}; \Omega_{\rm m,0}(w), w)=\chi(z_{\rm in}; \Omega^{\rm Planck}_{\rm m,0}, w=-1)
\eeq
where we have on the right hand side the standard $\Lambda$CDM value, evaluated by using the value $\Omega^{\rm Planck}_{\rm m,0}= \Omega^{\rm Planck}_{b,0}+\Omega^{\rm Planck}_{c,0}=0.02222 h^{-2}+ 0.1197 h^{-2}$, with $h=0.6731$,  which corresponds to the estimate deduced from the measurements by the Planck satellite \cite{Ade:2015xua}. We take $z_{\rm in}=100$, deep in the matter dominated era, when the effects of dark energy are negligible.

\subsection{The combination $c_s \DDt^{1/2}$}
\label{app2}

Here we provide details on the calculation of $c_s \alpha^{1/2}$.
For convenience we define the parameter
\be
\eta \equiv  \frac{1}{3} \left(3+ 2 \frac{\dot H}{H^2 } \right) =-\wDE \, \frac{(1-\Omega_{\rm m,0})a^{-3\wDE}}{\Omega_{\rm m,0}+(1-\Omega_{\rm m,0}) a^{-3\wDE}} \;, \label{eta_def}
\ee
which enters naturally in eqs.~\eqref{Omegab} and \eqref{Omegac}. For $\alphaM = \gammac=0$, the fraction that appears  on the right hand side reduces to  the energy density fraction of dark energy, $1-\Omega_{\rm m}$, but this is not the case in general. 
From eq.~\eqref{ss}, the combination $c_s^2 \DDt$ reads 
\be
c_s^2 \DDt = (1+\alphaB) (3 - 3\eta - 2 \xi) - 3 \Omega_{\rm m}-2\frac{\dot \alpha_\text{B} }{H}  \, , \label{csalpha}
\ee
where $\eta$ and $\xi$ are  defined above, respectively in eqs.~\eqref{eta_def} and \eqref{parameters}.
By using eqs.~\eqref{defparam} and \eqref{timedep_def} and  the background evolution equations \eqref{Omegab} and \eqref{Omegac} to evaluate $\dot \alpha_{\rm B}$ in this expression, 
this  can be written as
\be
\label{eqtheta}
\begin{split}
{c_s^2 \DDt} = 3 (1-\Omega_{\text{m}} -\eta) +   &  \alphaB  \bigg[ 1- 3 \eta \bigg(1 + 2  \frac{\Omega_{\rm m}}{1 - \Omega_{\text{m}}} \bigg)    -  2 ( \alphaM - 3  \gammac\,  \omega_c  )  \frac{\Omega_{\rm m}}{1-\Omega_{\text{m}}} \bigg] \\
&- 2 \alphaB^2 -2 \alphaT  \big(1+ \alphaB  \big)^2 + 2 \alphaM  ( 1 + \alphaB )     \, .
\end{split}
\ee
Finally, one can replace $\gammac$ by its expression \eqref{gamma_evol} given in terms of ${c_s \DDt^{1/2}}$.

The equation \eqref{eqtheta} is thus a quadratic equation for $X\equiv c_s\alpha^{1/2}$, of the form 
\be
\label{quad_eq}
X^2-BX -C=0,
\ee
where 
\be
B=\sqrt{2}  \frac{ \omega_c \, \Omega_{\rm m}}{1-\Omega_{\rm m}}\beta_\gamma \, \alphaB
\ee
and
\be
C=3(1-\Omega_{\rm m}-\eta)+\left[1-3\eta\frac{1+\Omega_{\rm m}}{1-\Omega_{\rm m}}-2 \frac{\Omega_{\rm m}}{1-\Omega_{\rm m}}\alphaM\right]\alphaB-2\alphaB^2-2\alphaT (1+\alphaB)^2+2\alphaM (1+\alphaB)\,.
\ee
Let us extract from this quadratic equation the relevant solution.

Let us start with the case $\alpha_{\text{B},0}=0$, which implies 
\be
B=0,\qquad C=3(1-\Omega_{\rm m}-\eta)-2(\alphaT-\alphaM) \qquad (\alpha_{\text{B},0}=0)
\ee
in  \eqref{quad_eq}, and the solution is therefore
\beq
c_s \alpha^{1/2}= \pm \sqrt{3(1-\Omega_{\rm m}-\eta)+2(\alphaM-\alphaT)}\,, \qquad (\alpha_{\text{B},0}=0)\,.
\eeq
Both signs of this solution can be chosen and lead to the same phenomenology as long as the sign of $\beta_\gamma$ is chosen to obtain the same $\gamma_c$.
In the matter dominated era, corresponding to $\Omega_{\rm m} \rightarrow 1$ and $\eta\rightarrow 0$, the stability condition thus imposes 
\beq
\alpha_{\text{M},0}\geq \alpha_{\text{T},0}\qquad (\alpha_{\text{B},0}=0)\,.
\eeq

Let us now consider the case $\alpha_{\text{B},0}\neq 0$. In the past limit $\Omega_{\rm m} \rightarrow 1$, one finds that 
\beq
B=\sqrt{2}\beta_\gamma \omega_c \frac{\Omega_{\rm m} \alphaB}{1-\Omega_{\rm m}}\quad \rightarrow\quad  \sqrt{2}\beta_\gamma \omega_c \frac{\alpha_{\text{B},0}}{1-\Omega_{\rm m,0}}
\eeq
behaves like a constant, while $C\rightarrow 0$. Consequently, the  two solutions of the quadratic equation in this limit are  $X=0$ and $X= \sqrt{2}\beta_\gamma \omega_c \alpha_{\text{B},0}/(1-\Omega_{\rm m,0})$. In order to recover a standard matter dominated regime with $\gamma \rightarrow 0$ in the past limit $\Omega_{\rm m}\rightarrow 1$, one needs to pick up the $X=0$ solution in the past. This determines the choice of the sign among  the two solutions 
\beq
X=\frac{B\pm\sqrt{B^2+4C}}{2}\,,
\eeq
which yield $X=(B\pm |B|)/2$ in the limit $\Omega_{\rm m} \rightarrow 1$. One thus concludes that, depending on the sign of $\alpha_{B,0}\, \beta_\gamma$, the solution is 
\begin{align}
c_s \DDt^{1/2} & =\frac{B-\sqrt{B^2+4C}}{2}\,, \qquad (\beta_\gamma \,\alpha_{\text{B},0}>0) \\
c_s \DDt^{1/2} & = \pm \sqrt{C} \;,\qquad (\beta_\gamma \,\alpha_{\text{B},0}=0)   \label{betazero}\\
{c_s \DDt^{1/2}} & =\frac{B+\sqrt{B^2+4C}}{2}\,. \qquad (\beta_\gamma \,\alpha_{\text{B},0}<0) 
\end{align}
As above, both signs on the right hand side of eq.~\eqref{betazero} can be chosen.
The stability condition $c_s^2 \alpha>0$  is obtained by requiring that the above solutions  are real.

\section{Matter evolution equations in a generic frame}
\label{app:dt}

For completeness, we provide here the evolution equations for matter in a generic frame. 
In a generic frame $g_{\mu \nu}$ where both baryons and CDM are nonminimally coupled, eqs.~\eqref{eqmatterb} and \eqref{eqmatterc} read
\begin{align}
\ddot \delta_b + (2 + 3 \gamma_b) H \dot \delta_b &= \frac32  H^2  \Omega_{\rm m}(1+ \Upsilon_b) \delta_{\rm m}  \;, \label{eqmatterbgf} \\
\ddot \delta_c + (2 + 3 \gammac) H \dot \delta_c &= \frac32  H^2 \Omega_{\rm m} (1+ \Upsilon_c) \delta_{\rm m}   \;, \label{eqmattercgf}
\end{align}
with 
\begin{align}
 \Upsilon_b&={\alpha}_{\text T}+ \beta_\xi^2+( \beta_{\gamma_b}^2+2 \beta_{\gamma_b} \beta_\xi){\omega}_b {b}_b+\left[ \beta_{\gamma_b} \beta_{\gamma_c}+ \beta_\xi( \beta_{\gamma_b}+ \beta_{\gamma_c})\right]{\omega}_c {b}_c \, ,\\
 \Upsilon_c&={\alpha}_{\text T}+ \beta_\xi^2+\left[ \beta_{\gamma_b} \beta_{\gamma_c}+\beta_\xi(\beta_{\gamma_b}+\beta_{\gamma_c})\right]{\omega}_b {b}_b+( \beta_{\gamma_c}^2+2 \beta_{\gamma_c} \beta_\xi){\omega}_c {b}_c \, .
\end{align}
For the case discussed in the main text of minimally coupled baryons, i.e.~$\beta_{\gamma_b}=0$,  one recovers the expressions in eq.~\eqref{Ups_g}.

Under a frame transformation \eqref{transf_cd}, $\omega_I=\Omega_I/\Omega_{\text m}$ does not change. Moreover, in the quasi-static limit the density contrasts $\delta_I$ does not change either (the explicit transformations are discussed in \cite{Gleyzes:2015pma}). In particular, this implies that $\tilde{b}_I=b_I$. Therefore, by using the transformations of the $\alpha$'s given in eq.~\eqref{alphatilde}, $\gamma_{I}$ given in eq.~\eqref{changegamma} and those of the $\beta$'s  given in eq.~\eqref{transbeta}, one finds the expressions for $\Upsilon_{b,c}$ in the frame $\tilde g_{\mu \nu}$,
\be
\label{changeUps}
\begin{split}
\tilde \Upsilon_b&= (1+\Upsilon_b)(1+\alphaD) -1 \, ,\\
\tilde \Upsilon_c&=(1+\Upsilon_c)(1+\alphaD) -1 \, .
\end{split}
\ee
Using the expressions above and that the factors $1+\alphaD $ is cancelled by the change of time between the two frames, $\text{d}\tilde{t}=\sqrt{ {C}/({1+\alphaD})}\text{d}t$, one can check that the form of eqs.~\eqref{eqmatterbgf} and \eqref{eqmattercgf} is frame-independent. 

\bibliographystyle{utphys}
\bibliography{EFT_DE_biblio2}

\end{document}